\documentclass[11pt,a4paper]{article} 
\usepackage{jheppub}

\usepackage{latexsym}
\usepackage{euscript}
\usepackage{booktabs}
\usepackage{latexsym} 	
\usepackage{amsmath}  	
\usepackage{amsbsy}
\usepackage{epsfig}     
\usepackage{graphicx}
\usepackage{latexsym}
\usepackage{euscript}
\usepackage{amsfonts}
\usepackage{mathtools} 
\usepackage{braket}
\usepackage{amsmath}
\usepackage{amssymb}
\usepackage{slashed}
\usepackage{float}
\usepackage{pgf}
\usepackage{dcolumn}

\numberwithin{equation}{section}

\newcommand{\Tr}{\textrm{Tr}}
\newcommand{\tr}{\textrm{Tr}}

\newcolumntype{.}{D{.}{.}{-1}}

\newcommand{\re}{\operatorname{Re}}

\newcommand{\retr}[1]{(\textrm{Re}\textrm{Tr})}

\newcommand{\bi}  {\begin{itemize}}
\newcommand{\ei}  {\end{itemize}}

\newcommand{\bc}  {\begin{center}}
\newcommand{\ec}  {\end{center}}

\newcommand{\beq}{\begin{equation}}
\newcommand{\eeq}{\end{equation}}
\newcommand{\bea}{\begin{eqnarray}}
\newcommand{\eea}{\end{eqnarray}}

\newcommand{\be}{\begin{equation}}
\newcommand{\ee}{\end{equation}}

\renewcommand{\bra}{\langle}
\renewcommand{\ket}{\rangle}

\newcommand{\om}{\omega}

\title{
Electrical conductivity and charge diffusion in thermal QCD from the lattice 
}

\author[a]{Gert Aarts,}
\author[a]{Chris Allton,} 
\author[a,b]{Alessandro Amato,}
\author[c]{Pietro Giudice,}
\author[a]{Simon Hands}
\author[d]{and Jon-Ivar Skullerud}

\affiliation[a]{Department of Physics, College of Science, Swansea University \\ Swansea SA2 8PP, United Kingdom}
\affiliation[b]{Department of Physics and Helsinki Institute of Physics \\ P.O.\ Box 64, FI-00014 University of Helsinki, Finland }
\affiliation[c]{Universit\"at M\"unster, Institut f\"ur Theoretische Physik \\ Wilhelm-Klemm-Str.~9, D-48149 M\"unster, Germany}
\affiliation[d]{Department of Mathematical Physics, National University of Ireland Maynooth \\ Maynooth, Co Kildare, Ireland}

\emailAdd{g.aarts@swan.ac.uk}
\emailAdd{ c.allton@swan.ac.uk}
\emailAdd{pyaa@swan.ac.uk}
\emailAdd{p.giudice@uni-muenster.de}
\emailAdd{s.hands@swan.ac.uk}
\emailAdd{jonivar@thphys.nuim.ie}

\abstract{
  We present a lattice QCD calculation of the charge diffusion
coefficient, the electrical conductivity and various susceptibilities
of conserved charges, for a range of temperatures below and above the
deconfinement crossover. The calculations include the contributions
from up, down and strange quarks.  We find that the
diffusion coefficient is of the order of $1/(2\pi T)$ and has a dip
around the crossover temperature.  Our results are obtained with
lattice simulations containing $2+1$ dynamical flavours on anisotropic
lattices. The Maximum Entropy Method is used to construct spectral
functions from correlators of the conserved vector current.
}


\begin{document}
\maketitle


\section{Introduction}

Recently, there has been a great deal of progress in the understanding
of the dynamical and static properties of the quark-gluon plasma
(QGP).  The current theoretical interpretation of heavy-ion collision
experiments at RHIC and CERN consists of a hydrodynamical description
of the evolution of the fireball, see e.g.\ the reviews \cite{Heinz:2013th,Gale:2013da}.
This relies on the
QGP medium thermalising after a very short time of less than 1 fm/c;
the subsequent evolution is then modelled by viscous hydrodynamics,
until hadronisation.

The input parameters in the hydrodynamic evolution equations are the equation of state and, for the nonideal case, 
transport coefficients, such as viscosities and conductivities. These
quantities capture the dynamics from the underlying theory and hence
are determined by QCD. A first-principles determination must face
the challenge of strong coupling: it is now widely accepted that
dynamics in the QGP is strongly coupled, typically expressed via the
statement that the ratio of the shear viscosity to entropy density
$\eta/s$ is close to the value obtained in models with holographic duals \cite{Kovtun:2004de}.

Besides the shear and bulk viscosities, there is an increasing interest in another transport coefficient, namely the electrical conductivity, due to the role it plays in heavy-ion phenomenology. For instance, the conductivity has recently been discussed in the context of  charge density fluctuations \cite{Ling:2013ksb} and the evolution of strong electromagnetic fields produced in noncentral collisions
 \cite{Tuchin:2013ie,McLerran:2013hla,Gursoy:2014aka,Zakharov:2014dia,Tuchin:2014iua,Satow:2014lia}. It 
has also been suggested that experimental information on conductivity can be extracted from flow parameters in heavy-ion collisions
\cite{Hirono:2012rt}.

Using linear response theory, transport coefficients can be related to current-current  spectral functions in thermal equilibrium, via the celebrated Kubo relations \cite{kubo,Kadanoff1963419}. This opens up the possibility to apply lattice QCD at finite temperature as the nonperturbative tool of choice, provided that the analytical continuation from Euclidean to real time, or from Euclidean correlators to spectral functions, can be carried out reliably  \cite{Aarts:2002cc}. In recent years several results have been obtained along these lines. Refs.~\cite{Meyer:2007ic,Meyer:2007dy} contain the first (theoretically controlled) results for the shear and bulk viscosity, see also the review \cite{Meyer:2011gj}.
The best-studied transport coefficient is however the electrical conductivity, since the corresponding
Euclidean correlator can be computed numerically with high
precision. The first results were obtained using the staggered fermion formulation
\cite{Gupta:2003zh,Aarts:2007wj} in quenched QCD. Since then the study of the systematic uncertainties and the extension to Wilson-type quarks have taken a central role, in quenched QCD \cite{Ding:2010ga,Kaczmarek:2013dya} and in dynamical QCD with $N_f=2$ and $2+1$ flavours \cite{Brandt:2012jc,Amato:2013naa}. All studies \cite{Aarts:2007wj,Ding:2010ga,Kaczmarek:2013dya,Brandt:2012jc,Amato:2013naa} are in qualitative agreement: around $T=1.5T_c$, 
where $T_c$ is the crossover temperature, the conductivity is of the order of $\sigma=0.2-0.4C_{\rm em}T_c$, where $C_{\rm em}$ is an electromagnetic prefactor depending on the quark charges (see Sec.~\ref{sec:cond} below).
Recent non-lattice studies include Refs.\ \cite{Cassing:2013iz,Finazzo:2013efa,Steinert:2013fza,Yin:2013kya,Lee:2014pwa,Puglisi:2014pda,Puglisi:2014sha,Greif:2014oia}.

The most detailed lattice study so far can be found in
Ref.~\cite{Amato:2013naa}, where the temperature dependence of
$\sigma/T$ across the deconfinement transition was studied for the
first time, over a range of temperatures corresponding to $0.63-1.9T_c$. In that analysis we used a QGP with $N_f=2+1$ flavours but only 
included the light quark contribution to the conserved vector current.
Here we improve upon those results by also including the strange quark contribution and comparing the relative contributions. Moreover we compute various susceptibilities, including the charge susceptibility $\chi_Q$, which allows us to compute for the first time the charge diffusion coefficient $D=\sigma/\chi_Q$ in a self-contained calculation. 
Since the diffusion coefficient can also be computed in strongly coupled theories that permit a holographic prescription, with the characteristic result that $D=1/(2\pi T)$ in ${\cal N} = 4$ Yang-Mills theory at nonzero temperature  \cite{Policastro:2002se,Teaney:2006nc,Son:2007vk}, this direct computation allows us to compare QCD with strongly coupled gauge theories which have a dual formulation.

The remainder of the paper is organised as follows. 
In the next section we start with the information on the lattice action and 
ensembles used in this work, followed by a determination of the 
crossover transition temperature using the renormalised Polyakov loop 
and the chiral susceptibility in Sec.~\ref{sec:Tc} and the baryon, 
isospin and charge susceptibilities in Sec.~\ref{sec:susc}. 
In Sec.~\ref{sec:cond} we turn to the electrical conductivity, and present 
the Euclidean correlators and their corresponding spectral functions 
determined via the Maximum Entropy Method. The systematics in this 
construction are discussed in some detail. Finally, our results for the 
diffusion coefficient, obtained using the results from the two preceding 
sections, are presented in Sec.~\ref{sec:diff}. 
We summarise our findings and provide a brief outlook in
Sec.~\ref{sec:conclude}.

\section{$N_f=2+1$ lattice details}
\label{sec:lattice}

In this work we follow the Hadron Spectrum Collaboration~\cite{Edwards:2008ja} 
and use a 
Symanzik-improved anisotropic gauge action with tree-level mean-field coefficients 
and a mean-field--improved Wilson-clover fermion action with stout-smeared links \cite{Morningstar:2003gk}. The anisotropy, with a reduced temporal lattice spacing $a_\tau<a_s$, is crucial to obtain a better resolution of the correlation functions, especially at higher temperatures.
This will be discussed further below.
Anisotropy introduces two new bare parameters in the action, the bare
gauge and fermion anisotropies, which are nonperturbatively tuned to
give the desired renormalised anisotropy, $\xi = a_s/a_\tau$,  common to the gauge and fermionic degrees of freedom.
The ensembles employed here are part of our ``2nd generation" data set \cite{Allton:2014uia} and were previously used for a determination of the conductivity (from two light flavours only) \cite{Amato:2013naa} and the bottomonium spectrum at nonzero temperature \cite{Aarts:2014cda}.

The gauge action takes the form
\be
S_G = \frac{\beta}{\gamma_g}
\sum_{x,i\neq i^\prime} \left[
    \frac{5}{6 u_s^4     }{\cal P}_{ii^\prime}(x)
  - \frac{1}{12 u_s^6    }{\cal R}_{ii^\prime}(x)\right]
+ \beta \gamma_g \sum_{x,i} \left[
    \frac{4}{ 3 u_s^2 u_\tau^2}{\cal P}_{i4}(x)
  - \frac{1}{12 u_s^4 u_\tau^2}{\cal R}_{i4}(x)\right],
\label{eq:sg}
\ee
where ${\cal P}$ and ${\cal R}$ are the $1\times1$ plaquette and
$2\times1$ rectangular Wilson loops, $u_{s(\tau)}$ are the spatial
(temporal) mean links, $\gamma_g$ is the
bare gauge anisotropy and, as usual, $\beta=2 N_c/g^2$ with
$N_c=3$ colours.

The fermion action (for a single flavour) reads 
\begin{align}
S_F = \sum_{x} \overline{\psi }(x) \bigg[ &
 \hat m_0 +  \gamma_4 \nabla_4 - \frac{1}{2} \nabla_4^2
 +\frac{1}{\gamma_f} \sum_i \left( \gamma_i    \nabla_i   + \frac{1}{2}  \nabla_i^2 \right)
 \nonumber \\
 & - \frac{1}{2}  c_\tau \sum_i     \sigma_{4 i}    F_{4i}
  -\frac{c_s}{2\gamma_g}    \sum_{i<i^\prime} \sigma_{ii^\prime} F_{ii^\prime}
\bigg] \psi (x),
\label{eq:sf}
\end{align}
where $\hat m_0=a_\tau m_0$ is the bare fermion mass,  $\gamma_f$ the bare fermion anisotropy, 
$\nabla_\mu$ covariant finite differences,
$\sigma_{\mu\nu} = \frac{i}{2}[\gamma_\mu, \gamma_\nu]$, 
and the clover coefficients
\begin{equation}
c_\tau = \frac{1}{2} \left(\frac{\gamma_g}{\gamma_f}+\frac{1}{\xi}\right)
\frac{1}{\tilde{u}_s^2}\;,
\quad\quad\quad
c_s    = \frac{\gamma_g}{\gamma_f} \frac{1}{ \tilde{u}_s^3}.
\end{equation}
The spatial gauge links in the fermion action have been stout smeared
\cite{Morningstar:2003gk} with smearing weight $\rho=0.14$ and
$n_\rho=2$ iterations, and $\tilde{u}_{s}$ is the mean value of the
spatial stout-smeared links.

\begin{table}[t]
\begin{center}
\begin{tabular}{ll. | l}
       \hline\hline
 gauge coupling &  $\beta$ 		& 1.5  
 &   \\
 bare gauge anisotropy	 & $\gamma_g$ 	& 4.3  
 & $a_s$ = $0.1227(8)$ fm      \\
 bare fermion anisotropy & $\gamma_f$  	& 3.4
 & $a_\tau^{-1}=5.63(4)$ GeV     \\
 spatial clover coefficient & $c_s$	& 1.5893
 & $\xi=a_s/a_\tau$ = 3.5 \\      	 
 temporal clover coefficient & $c_\tau$	& 0.9027
& $M_\pi = 384(4)$ MeV \\
bare light quark mass	& $\hat m_{ud}$ &  -0.0840
& $M_\pi/M_\rho = 0.446(3)$ \\
 bare strange quark mass & $\hat m_{s}$ &  -0.0743
& \\ 
 	   \hline\hline
    \end{tabular}
\caption{Lattice parameters.}
\label{tab:lattice-a}
\end{center}
\end{table}

The choice of bare parameters is given in  Table~\ref{tab:lattice-a}
and follows the tuning by the Hadron Spectrum Collaboration
\cite{Edwards:2008ja}. The resulting renormalised anisotropy is
$\xi=3.5$. The two degenerate light quarks yield $M_\pi=384(4)$ MeV
(2.8 times larger than the physical pion),
corresponding to $M_\pi/M_\rho = 0.446(3)$~\cite{Lin:2008pr}, while
the third flavour is tuned to the strange quark
mass~\cite{Edwards:2008ja}.

\begin{table}[t]
\begin{center}
\begin{tabular}{cccccccc}
    \hline\hline
      \multicolumn{2}{c}{$N_s$} & $N_\tau$  & $T$  [MeV] & $T/T_c$  & $N_{\texttt{CFG}}$ & $N_{\texttt{SRC}}$  \\
\hline\hline
   24 & 32 & 16 & 352 & 1.90 & 1059  &4 \\ 
   24 &       & 20 & 281 & 1.52 & 1001  &4 \\ 
   24 & 32 & 24 & 235 & 1.27 & 500   &4 \\ 
   24 & 32 & 28 & 201 & 1.09 & 502   &4 \\ 
   24 & 32 & 32 & 176 & 0.95 & 501   &4 \\ 
   24 &       & 36 & 156 & 0.84 & 501   &4 \\
   24 & 	& 40 & 141 & 0.76 & 523   &4 \\
     	& 32 & 48 & 117 & 0.63 & 601   &4 \\
   24	&	&128& 44	  & 0.24 & 401	&1 \\
   \hline\hline
   \end{tabular}
\caption{Details of the ensembles. The lattice size is $N_s^3\times N_\tau$, with the temperature $T=1/(a_\tau N_\tau)$. 
  $N_{\texttt{CFG}}$ ($N_{\texttt{SRC}}$) denote the number of configurations
  at each volume (the number of source positions within the volume) used for the analysis of the conductivity.
  }
\label{tab:lattice-b}
\end{center}
\end{table}

Details of the finite-temperature ensembles are given in Table~\ref{tab:lattice-b}. Note that there are five ensembles in the hadronic phase and four in the quark-gluon plasma phase. The determination of the pseudo-critical temperature $T_c$ is discussed in the next section.
In order to look for finite-size effects, we have generated configurations with two spatial volumes at four different temperatures, with spatial 
extents of $\sim 2.9$ respectively 3.9 fm ($N_s=24$ and $32$).
Access to the zero-temperature configurations ($N_\tau=128$) has been kindly provided to us by the Hadron Spectrum Collaboration.
The ensembles were generated using the Rational Hybrid Monte Carlo
(RHMC) algorithm with multiple timescale integration and Hasenbusch
preconditioning for the light quarks, using the Chroma
software suite \cite{Edwards:2004sx} with Bagel routines
\cite{Boyle:2009vp}.  For further details about the algorithm, we
refer to Refs.~\cite{Edwards:2008ja,Lin:2008pr}.  After 1000
thermalisation trajectories (2000 for the $32^3\times24$ ensemble),
configurations were 
sampled every 10 RHMC trajectories, except for the $32^3\times48$
ensemble where the sampling frequency was every 5 trajectories.  The
plaquette and Polyakov loop autocorrelation times were found to be
between 2 and 30 trajectories.

\section{Deconfinement and chiral transition}
\label{sec:Tc}

After a determination of the lattice spacing, the temperature can be specified in MeV very precisely using the standard relation $T=1/(a_\tau N_\tau)$. However, it is desirable to express the temperature in units of $T_c$, the crossover temperature, especially since the two light quarks are heavier than in Nature  (for a lattice study of the transition with physical quark masses, see e.g.\ Ref.\ \cite{Aoki:2006we}).  In order to do so, we use the renormalised Polyakov loop as an indicator of the deconfinement transition, following closely the renormalisation procedure described in Ref.\ \cite{Borsanyi:2012xf}.

The Polyakov loop expectation value $L$ is related to the free energy of a static quark $F$ via
\be
L(T) = e^{-F(T)/T}.
\end{equation}
However, $F$ is only defined up to an additive renormalisation constant  $\Delta F$, 
which depends on the gauge coupling and other bare parameters but not on the temperature. Expressing the renormalised free energy $F_R$ as $F_R=F+\Delta F$ allows us write the renormalised Polyakov loop as
\be
L_R(T) \equiv e^{-F_R(T)/T} = e^{-(F(T)+\Delta F)/T} = Z_L^{N_\tau} L(T),
\ee
which defines the multiplicative renormalisation constant $Z_L$.
Following Ref.\ \cite{Borsanyi:2012xf}, we impose a renormalisation condition at a reference temperature $T_R$, by requiring that
\begin{equation}
L_R(T_R) \equiv \mbox{constant},
\label{eq:lr}
\end{equation}
which fixes $Z_L$.

\begin{figure}[t]
\begin{center}
\includegraphics[width=\textwidth]{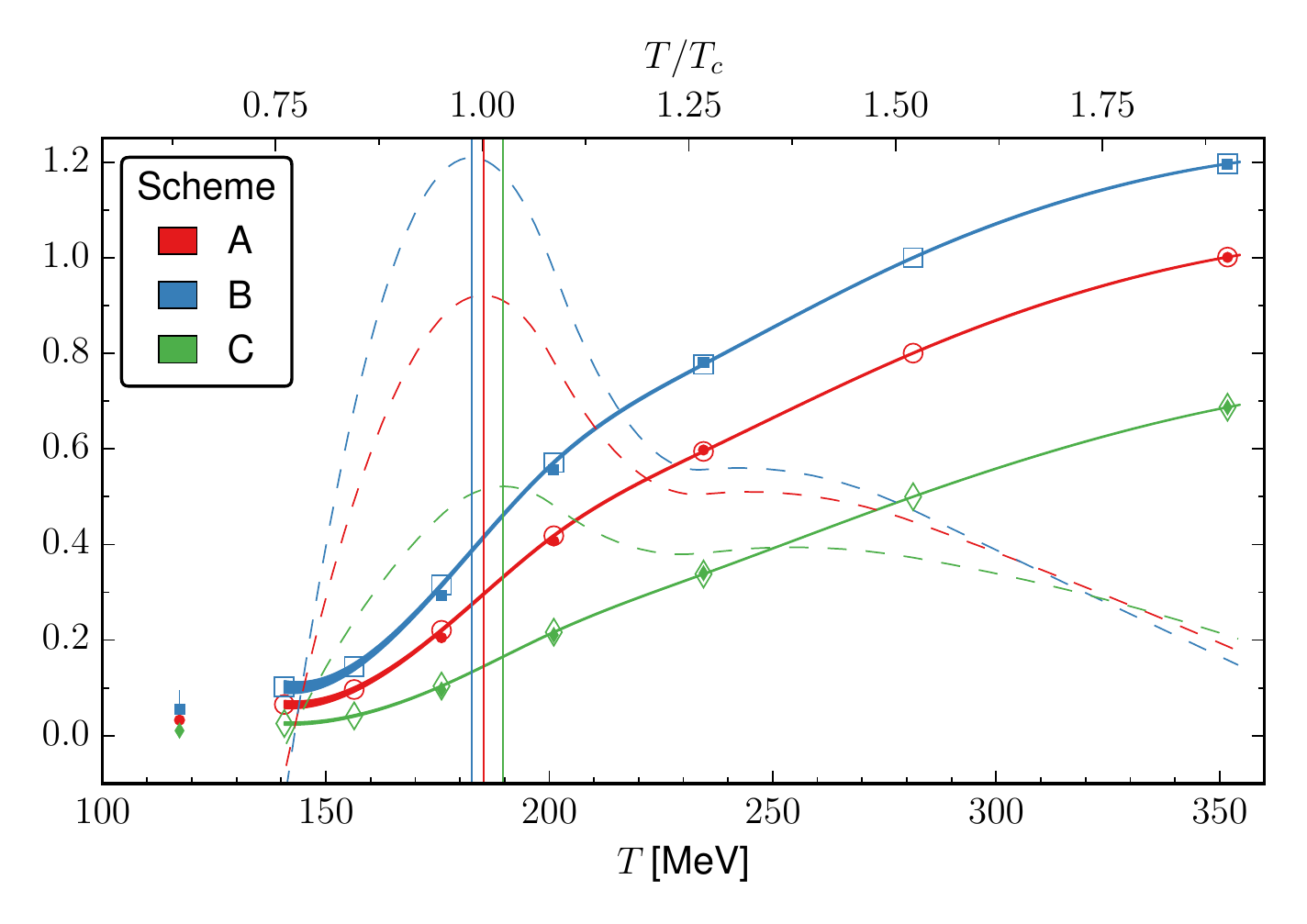}
\caption{Renormalised Polyakov loop $L_R$ as a function of temperature 
as defined by the procedure explained in the text.  
Three renormalisation schemes are considered, A: $L_R(N_\tau=16) = 1.0$, B: $L_R(N_\tau=20) = 1.0$ and C: $L_R(N_\tau=20) = 0.5$.
Solid (open) symbols use a spatial volume of $32^3$ ($24^3$).
The data points are connected with curves obtained by cubic splines; their derivatives, depicted by the dashed curves, represent the Polyakov loop susceptibility. The vertical lines indicate the peak of the susceptibility in each scheme.
  }
\label{fig:poly}
\end{center}
\end{figure}

Figure~\ref{fig:poly} shows the Polyakov loop with three different
renormalisation schemes corresponding to different choices of $T_R$
and the constant in Eq.~(\ref{eq:lr}), as detailed in the figure
caption. The data are interpolated using cubic splines, with the
  statistical uncertainty given by the thickness of the three
  interpolating curves; it can be seen to be negligible.  We then
  obtain the Polyakov loop susceptibility as the derivatives of the
  interpolating curves for the Polyakov loop in each of the three
  schemes.  The peak positions are indicated with the vertical lines
  in Fig.~\ref{fig:poly} and give us the point of inflection at
  $N_\tau^{\rm crit} = 30.4(7)$ or $1/N_\tau^{\rm crit} = a_\tau T_c =
  0.0329(7)$, where the error reflects the systematic error coming
  from the spread of the three renormalisation schemes.  This
  corresponds to a deconfinement critical temperature of $T_c =
  185(4)$ MeV.  We note that neither chiral nor continuum
extrapolations have been performed in our analysis.

In the limit of massless quarks, QCD becomes classically invariant
under chiral transformations. This symmetry is spontaneously broken
at low temperature.
For physical masses, even if the chiral
symmetry is explicitly broken, the associated order parameter shows a clear
transition signal at a certain temperature. 
The chiral condensate and, in particular, its
susceptibility $\chi_c$ are commonly used to define 
the crossover transition temperature. 
We have determined the chiral susceptibility due to the two light flavours, using  \cite{Aoki:2006br,Borsanyi:2010bp,Bazavov:2011nk}
\bea
\nonumber
\chi_c &=& \frac{T}{V} \frac{\partial^2 \ln Z}{\partial m^2} =
 \chi_{\rm disc} + \chi_{\rm conn}, \\
 \label{chiraleqs}
 \chi_{\rm disc} &=& \frac{4T}{V} \left[ 
  \left\bra \left(\Tr M^{-1} \right)^2 \right\ket -
\left\bra \Tr M^{-1} \right\ket^2 \right],\\
\nonumber
\chi_{\rm conn} &=& - \frac{2T}{V} \left\bra \Tr M^{-1} M^{-1} \right\ket, 
\eea
where $Z$ is the partition function, $M$ the fermion matrix, 
$V$ the spatial volume and $m$ the degenerate light quark mass.
Moreover, we introduce here the connected $\chi_{\rm conn}$ and the disconnected
$\chi_{\rm disc}$ contributions to the susceptibility.
The traces in Eq.~(\ref{chiraleqs}) are determined using 16 noise vectors
for the disconnected contribution and 4 for the connected one. 
Because we change the temperature by changing the value of $N_\tau$ rather than the lattice spacing, the (additive and multiplicative) renormalisation of $\chi_c$ is the same for all temperatures. 
A peak in the susceptibility occurs therefore at the same temperature for the
  renormalised and unrenormalised $\chi_c$. 
We hence show the bare susceptibility and are only interested in the overall shape, rather than the absolute value.

\begin{figure}[t]
\begin{center}
\includegraphics[width=\textwidth]{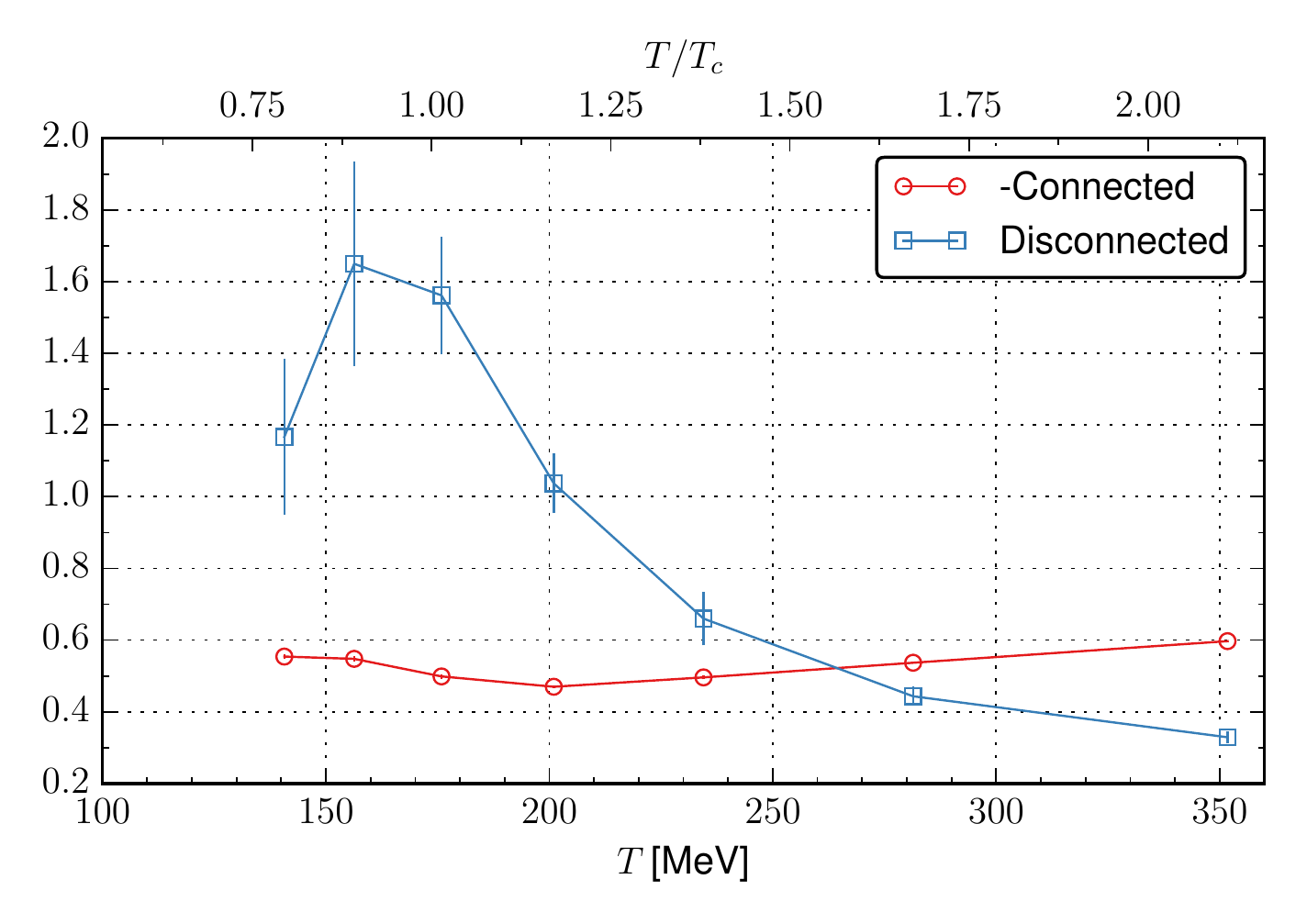}
\caption{Disconnected and minus the connected contributions 
to the (unrenormalised) chiral susceptibility $\chi_c$, computed on the $24^3\times N_\tau$ ensembles.
}
\label{fig:chiralsusc}
\end{center}
\end{figure}

The results are plotted in Fig.~\ref{fig:chiralsusc}. The connected 
contribution is not singular while the disconnected contribution shows
a peak.
 Because the determination of the disconnected term is particularly expensive at low temperature,  we could not determine the chiral critical temperature very precisely. Our best estimate is $T_c^{\chi}
  \approx 170(20)$ MeV, which is somewhat lower than the value
  obtained from the Polyakov loop. 
 In the following we will use the value for $T_c$ determined from
 the Polyakov loop as this has been obtained with a better precision.

\section{Susceptibilities}
\label{sec:susc}

Fluctuations of conserved charges are sensitive probes both of the
thermal state of the medium and of its critical behaviour.  They are
quantified by susceptibilities, defined as second (and higher)
derivatives of the free energy with respect to the chemical potential
associated with the investigated charge.  In QCD, assuming three
active light flavours, charges that can be studied include baryon number, electric charge and strangeness.  Their
susceptibilities probe the actual degrees of freedom that carry such
charges, i.e.\ quarks or hadrons.
Experimentally, fluctuations can be used to probe quark
confinement~\cite{Asakawa:2000wh} by studying event-by-event
fluctuations of charged particle ratios~\cite{Bleicher:2000ek}.
Susceptibilities show a rapid rise in the crossover region: at low
temperature they are small since quarks are confined; at high
temperature they are larger and they approach the ideal gas limit.
They have been studied by many groups in the past
\cite{Gottlieb:1988cq,Allton:2002zi,Gavai:2005sd, Bernard:2007nm, Borsanyi:2011sw, Bazavov:2013uja}.
Notably, so far, lattice studies have mainly employed staggered fermions. Here instead we use clover-improved Wilson fermions. 
For an earlier study using Wilson fermions see Ref.~\cite{Borsanyi:2012uq}.
The charge diffusion coefficient $D$ and the electrical conductivity $\sigma$ are related via the well-known relation $D=\sigma/\chi_Q$, where $\chi_Q$ is the charge susceptibility \cite{Kadanoff1963419}. In this section we determine $\chi_Q$ and various other (second-order) susceptibilities, defined as second
derivatives of the free energy with respect to the chemical potential associated with a conserved charge.

We introduce the quark number density and the quark number susceptibilities 
\begin{equation}
 \label{chi_ij}
 n_f = \frac{T}{V} \frac{\partial \ln{Z}}{\partial \mu_f},
 \quad\quad\quad\quad
 \chi_{ff'} = \frac{T}{V} \frac{\partial^2 \ln{Z}}{\partial \mu_f\partial \mu_{f'} } =  \frac{\partial n_f}{\partial \mu_{f'}},
\end{equation} 
where $Z$ is the partition function, $V$ the spatial volume,  and $\mu_f$ the quark 
chemical potentials for flavours $f \in \{ u,d,s \}$. 
Baryon ($B$), isospin ($I$) and electrical charge ($Q$) chemical potentials are related to the quark chemical potentials as
\be
\mu_u= \frac{1}{3}\mu_B+\frac{2e}{3}\mu_Q+\frac{1}{2}\mu_I,
\quad\quad
\mu_d= \frac{1}{3}\mu_B-\frac{e}{3}\mu_Q-\frac{1}{2}\mu_I,
\quad\quad
\mu_s= \frac{1}{3}\mu_B-\frac{e}{3}\mu_Q.
\ee
In general we denote the electrical charge of the quark as $eq_f$, with $e$ the elementary charge and $q_f=2/3$ or  $-1/3$ its fractional charge.

All desired quantities can now be expressed in terms of $n_f$ and $\chi_{ff'}$.
The baryon number density and baryon number susceptibility are given by
\be
n_B = \frac{T}{V} \frac{\partial \ln{Z}}{\partial \mu_B} = \frac{1}{3}\sum_f  n_f,
\quad\quad\quad
 \chi_B = \frac{\partial n_B}{\partial \mu_B}=
\frac{1}{9}\sum_{f,f'} \chi_{ff'},
\label{eq:B}
\eeq
the isospin density  and its susceptibility are given by
\be
n_I =  \frac{T}{V} \frac{\partial \ln{Z}}{\partial \mu_I} = \frac{1}{2}\left(n_u-n_d\right),
\quad\quad\quad
\chi_I = \frac{\partial n_I}{\partial \mu_I}
=
\frac{1}{4}\left( \chi_{uu} + \chi_{dd} - 2\chi_{ud} \right),
\label{eq:I}
\ee
and finally the electric charge density and its susceptibility are
\be
n_Q = \frac{T}{V} \frac{\partial \ln{Z}}{\partial \mu_Q} = e\sum_f  q_f n_f,
\quad\quad
 \chi_Q = \frac{\partial n_Q}{\partial \mu_Q} =  e^2\sum_{f,f'} q_f q_{f'} \chi_{ff'}.
\label{eq:Q}
\ee

To proceed we denote the fermion matrix on the lattice with $M$ and introduce the quark chemical potential $\mu_f$ in the usual way, i.e.\ as a constant imaginary vector potential in the temporal direction \cite{Hasenfratz:1983ba}. We then encounter the following derivatives \cite{Giudice:2013fza}
\begin{align}
& R_f^{(1)} = 
\frac{T}{V} \left\langle\tr{\left[ M^{-1} \frac{\partial M}{\partial \mu_f}  \right]}\right\rangle,
&& 
R_{ff'}^{(3)} = \frac{T}{V} \left\langle  \tr{\left[ M^{-1} \frac{\partial M}{\partial \mu_f} 
\right]}  \tr{\left[ M^{-1} \frac{\partial M}{\partial \mu_{f'}} 
\right]} \right\rangle,
\notag \\
& R_f^{(2)} = 
\frac{T}{V} \left\langle  \tr{\left[ M^{-1} \frac{\partial^2 M}{\partial \mu_f^2} 
\right]} \right\rangle,
&& R_f^{(4)} =
\frac{T}{V} \left\langle \tr{\left[ M^{-1} \frac{\partial M}{\partial \mu_f} 
M^{-1} \frac{\partial M}{\partial \mu_f} \right]}\right\rangle,
\label{terms}
\end{align}
where all expectation values are evaluated at vanishing chemical
potentials, $\mu_f=0$. 
It follows from symmetry that $n_f = R_f^{(1)}=0$. The diagonal and off-diagional susceptibilities are then written as
\be
\chi_{ff} = R_{ff}^{(3)} + R_f^{(2)}  - R_f^{(4)}, 
\quad\quad\quad
\chi_{ff'} = R_{ff'}^{(3)} \quad\quad (f \neq f'),
\ee
where we used the fact that the fermion matrix is the direct product of the fermion matrices for each flavour.
Denoting the degenerate light quarks with $\ell=u=d$, we find finally that
\begin{align} 
\notag
9\chi_B = &\; 4 R^{(3)}_{\ell\ell} + R^{(3)}_{ss} + 4R^{(3)}_{\ell s} + 2 \left( R^{(2)}_\ell -R^{(4)}_\ell \right)  + R^{(2)}_s - R^{(4)}_s,  \\
\notag
(9/e^2) \chi_Q= &\; R^{(3)}_{\ell\ell} + R^{(3)}_{ss} - 2 R^{(3)}_{\ell s} + 5 \left( R^{(2)}_\ell -R^{(4)}_\ell \right)  + R^{(2)}_s - R^{(4)}_s,  \\
2\chi_I = &\; R^{(2)}_\ell -R^{(4)}_\ell.
\label{eq:comb}
\end{align}
We note that for two degenerate light flavours the isospin
susceptibility $\chi_I$ does not depend on the disconnected
term $R^{(3)}$, while this term contributes more strongly
  to the baryon susceptibility $\chi_B$ than to the charge
  susceptibility $\chi_Q$.  Note that for
three degenerate flavours, $\chi_Q$ is also independent of $R^{(3)}$, since 
$R^{(3)}_{\ell\ell}= R^{(3)}_{\ell s}= R^{(3)}_{ss}$. 
The disconnected term is numerically the most expensive quantity
to be computed and it dominates the uncertainty of the final results.

\begin{figure}[t]
\begin{center}
\includegraphics{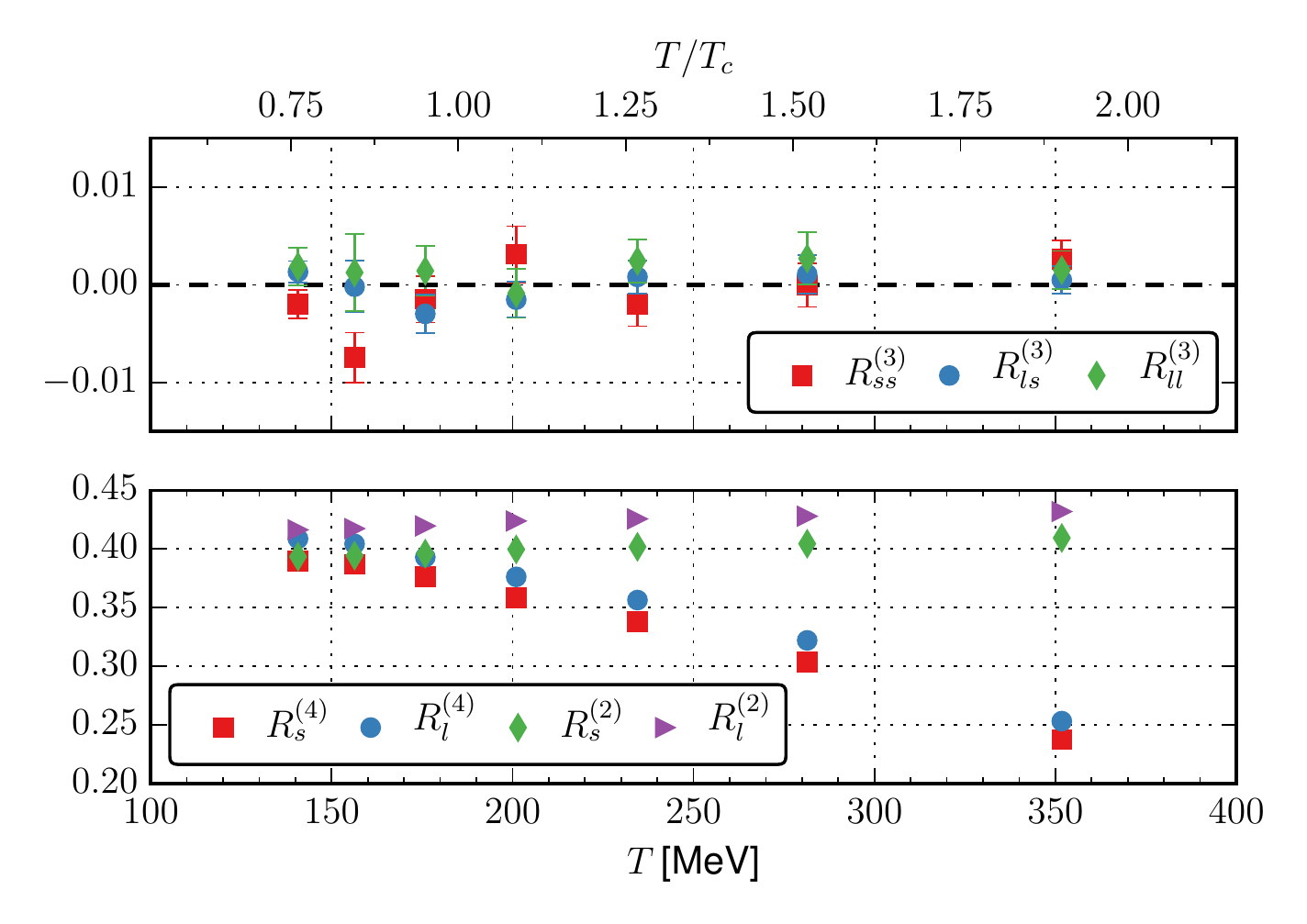}
     \caption{
    Disconnected ($R^{(3)}$, top)  and connected ($R^{(2,4)}$, below) contributions to the susceptibilities, for the light ($\ell$) and strange ($s$) quarks, versus temperature.
       }
\label{fig:susc1}
\end{center}
\end{figure}

We have determined the susceptibilities numerically on our $N_s=24$ ensembles, 
see Table~\ref{tab:lattice-a}. The traces in Eq.~(\ref{terms}) are
estimated stochastically, using $N_v=9$ noise vectors for the connected terms $R^{(2,4)}$. For the disconnected
term $R^{(3)}$, we use $N_v=200$
noise vectors in the $N_\tau=40$ case and $N_v=100$ at the other temperatures.
More technical details can be found in Ref.~\cite{Giudice:2013fza}.

\begin{figure}[t]
\begin{center}
\includegraphics{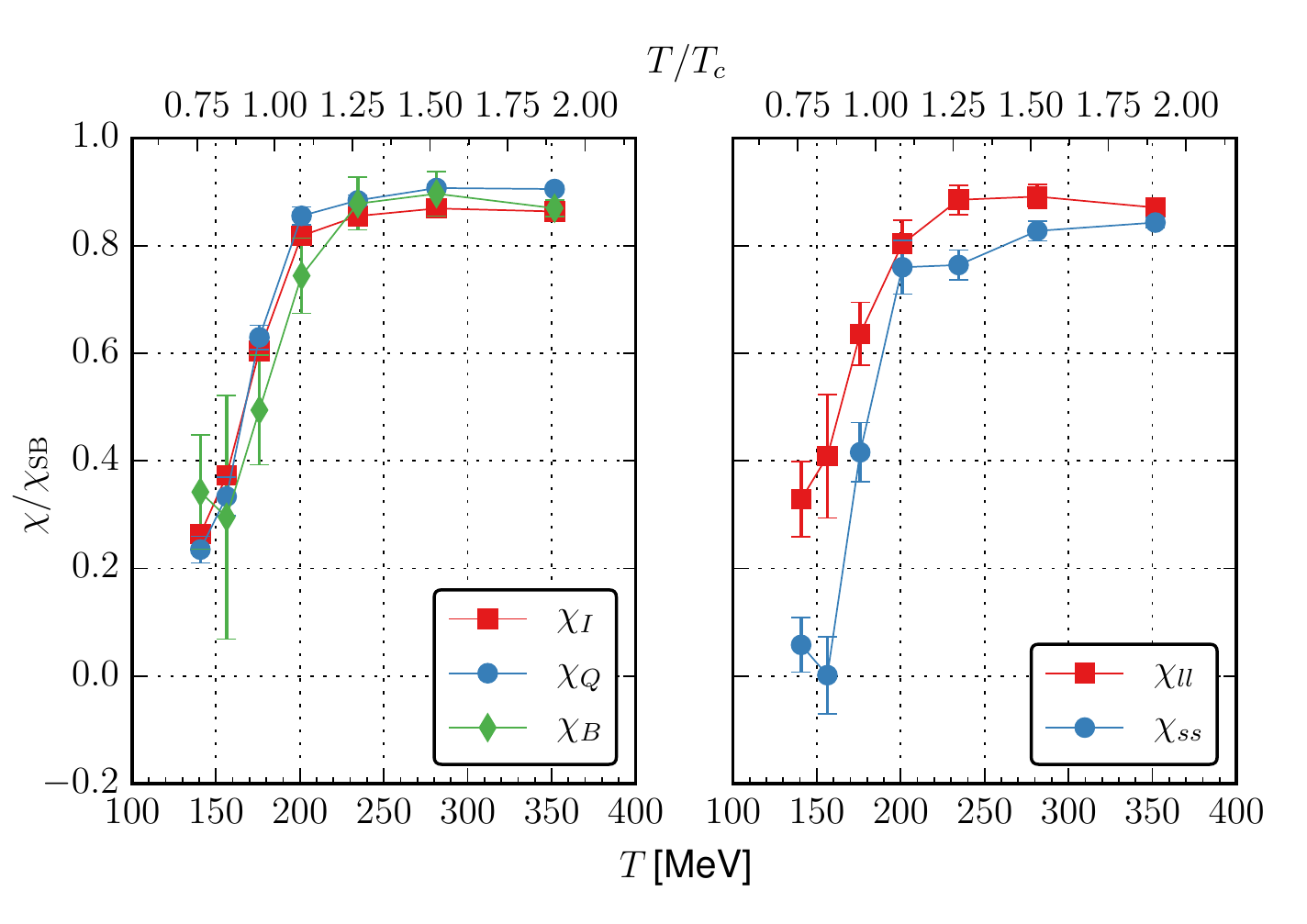}
    \caption{
       Isospin, charge and baryon number susceptibilities (left) and  quark number susceptibility for light and strange quarks (right), normalised with the corresponding quantity for free lattice fermions, denoted as $\chi_{\rm SB}$.
           }
\label{fig:susc2}
\end{center}
\end{figure}

In Fig.~\ref{fig:susc1}  we present the results for $R^{(2,3,4)}$, for both light and strange quarks. We observe that at high temperature, the dominant contribution comes from $R^{(2)}$ and  that  $R^{(3)}$ is compatible with zero within errors, for both the diagonal and off-diagonal components.  In this context we note that in  hard thermal loop (HTL) perturbation
theory~\cite{Blaizot:2001vr} the off-diagonal susceptibility is nonzero, showing a
clear correlation between different flavours. Also recent
lattice calculations~\cite{Borsanyi:2011sw} have shown a clear dip for
the off-diagonal term in the crossover region.  
Our result might be due to the relatively heavy sea quark masses.

The various susceptibilities $\chi_I$, $\chi_Q$ and $\chi_B$ are presented in Fig.~\ref{fig:susc2} (left), where all observables are normalised with the corresponding quantities for free lattice fermions, with the same lattice geometry 
\cite{Giudice:2013fza}. 
In the free case the bare fermion anisotropy is set equal to the 
renormalised value for our ensembles, while the bare quark mass is set to zero.
We have evaluated the effect of the uncertainty in the determination of 
$\xi$, see Ref.~\cite{Dudek:2012xn}, on our final results and found it to be a systematic effect of the order of $5 \%$.
In Fig.~\ref{fig:susc2} we clearly see a steep increase above $150$ MeV, and for $T \gtrsim 250$ MeV the value of the susceptibilities is around $85\%$ of the Stefan-Boltzmann value, i.e.\ the free case.
The result for $\chi_Q$ will be used in Sec.~\ref{sec:diff} to determine the diffusion coefficient.
The baryon number susceptibility behaves qualitatively in a similar way to the other two, but has larger errors due to the way the various terms combine, see Eq.\ (\ref{eq:comb}).
Our results appear consistent with the findings of other lattice groups \cite{Borsanyi:2011sw,Bazavov:2013uja} and also with resummed perturbation theory \cite{Andersen:2012wr}, in particular concerning the deviation from unity at the highest temperatures.

Finally, in Fig.~\ref{fig:susc2} (right) we show separately the quark number susceptibilities for light and strange
quarks, again normalised with the  corresponding quantity for free lattice fermions.  We see some indication for ``flavour separation'' during the QCD crossover transition, as discussed in Ref.~\cite{Ratti:2011au} and reported in Ref.~\cite{Bellwied:2013cta},
where a continuum extrapolated lattice QCD calculation was
performed (see, however, Ref.~\cite{Bazavov:2013dta}).

\section{Conserved vector currents and conductivity}
\label{sec:cond}

We consider the electromagnetic current for three flavours, 
\begin{equation}
\label{eq:emcur}
j^{\rm em}_\mu = \sum_f (eq_f) j_\mu^f
 = \frac{2e}{3}j^u_\mu - \frac{e}{3}j^d_\mu - \frac{e}{3}j^s_\mu\,,
\end{equation} 
where $j_\mu^f$ are the vector currents for each flavour and $eq_f$
are the corresponding electric charges.  
The Euclidean current-current correlator $G^{\rm em}_{\mu\nu}(\tau)$  is then
defined, at zero spatial momentum,  as
\begin{equation}
\label{eq:emcurcorr}
G^{\rm em}_{\mu\nu}(\tau) =  \int d^3x\, 
 \left\bra j^{\rm em}_\mu(\tau, {\mathbf x}) j^{\rm em}_\nu(0, {\mathbf 0})^\dagger\right\ket.
\end{equation}
This correlator admits a spectral representation of the form
\begin{equation}
\label{eq:spectr} 
 G^{\rm em}_{\mu\nu}(\tau) =  \int_0^\infty \frac{d\omega}{2\pi}\,
 K(\tau,\omega) \rho^{\rm em}_{\mu\nu}(\omega),
\end{equation}
with the kernel
\begin{equation}
 K(\tau,\omega)=
 \frac{\cosh[\omega(\tau-1/2T)]}{\sinh[\omega/2T]},
\end{equation}
where $\rho^{\rm em}_{\mu\nu}(\om)$ is the spectral function.

Application of linear response theory \cite{Kadanoff1963419} yields the Kubo formula for
the electrical conductivity $\sigma$, 
\begin{equation}
\label{eq:kubo}
  \frac{\sigma}{T} = \frac{1}{6T}  \lim_{\omega \rightarrow 0}\frac{\rho^{\rm
em}(\omega)}{\omega}, 
 \qquad\quad
 \rho^{\rm em}(\omega) = \sum_{i=1}^3 \rho^{\rm em}_{ii}(\omega)\;,
\end{equation}
where the spectral function $\rho^{\rm em}(\omega)$ has to be obtained
from the Euclidean correlator   
$G^{\rm em}(\tau) =\sum_i G_{ii}^{\rm em}(\tau)$ by
inverting Eq.~(\ref{eq:spectr}). 

It will be useful to normalise the electromagnetic observables by the sum
of the square of the individual quark charges, 
\begin{equation}
\label{eq:Cem}
C_{\rm em} = e^2 \sum_f  q_f^2,
\end{equation} 
which equals $2e^2/3$ for three flavours.  We then define
\begin{equation}
 G^{\rm em}(\tau) = C_{\rm em}\,G(\tau),
 \qquad\quad
 \rho^{\rm em}(\omega) = C_{\rm em}\,\rho(\omega),
\end{equation}
and consider $G(\tau)$ and $\rho(\om)$ from now on.  Where the light/strange quark contributions are shown separately, the corresponding correlators and spectral functions are normalised with the electromagnetic prefactor for two light quarks/one strange quark respectively.

\subsection{Correlators}
\label{subsec:cc}

\begin{figure}[t]
\begin{center}
\includegraphics[width=\textwidth]{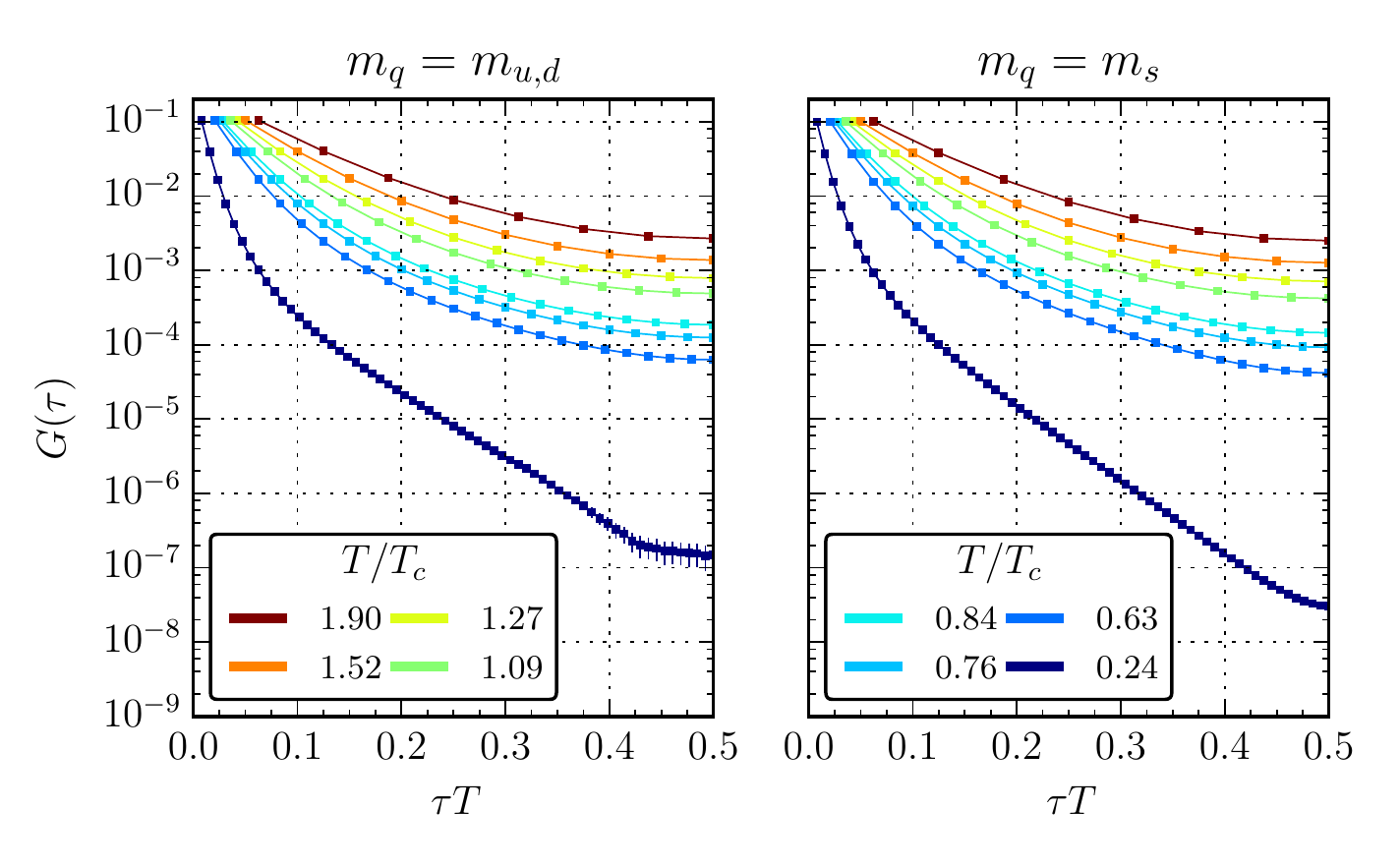}
\caption{
Conserved vector current correlator $G(\tau)$ in lattice units as a function of euclidean
time $\tau T$ at different temperatures for light (left) and
strange (right) quarks.
  }
\label{fig:consclover}
\end{center}
\end{figure}

We use the exactly conserved vector current on the lattice as an
interpolator for $j^{\rm em}_\mu$, since it is protected from
renormalisation under quantum corrections. It is defined as
\begin{equation}
V_{\mu}^{\text{C}}(x) = c_\mu \left[  \bar\psi(x+\hat\mu)(1+\gamma_\mu) 
U_\mu^\dagger(x) \psi(x)
  - \bar\psi(x) (1-\gamma_\mu) U_\mu(x) \psi(x+\hat\mu)\right],  
\end{equation}
where $c_4=1/2$, $c_i=1/(2\gamma_f)$ and $U_\mu(x)$ are the gauge links.
The two connected diagrams that contribute to the correlator
(\ref{eq:emcurcorr}) are 
\begin{align}
 \left\bra V_{\mu }^\text{C}(x)\,V_{\nu}^\text{C}(y)^\dagger \right\ket =   2c_\mu c_\nu 
&\,\re\,\tr \Big[  S(y+\hat\nu,x+\hat\mu)U_\mu^\dagger(x)\Gamma_\mu^+ S(x,y)U_\nu(y)\widetilde{\Gamma}_\nu^+
\nonumber\\
& -S(y,x+\hat\mu)U_\mu^\dagger(x)\Gamma_\mu^+ S(x,y+\hat\nu)U_\nu^\dagger(n)\widetilde\Gamma_\nu^-   \Big],
\end{align} 
 where $S(x,y)=\braket{\psi(x)\bar\psi(y)}$ is the fermion propagator, $\Gamma_\mu^\pm=1\pm\gamma_\mu$,
 $\widetilde\Gamma_\mu^\pm=1\pm\widetilde\gamma_\mu$, and $\widetilde\gamma_\mu=\gamma_4\gamma_\mu\gamma_4$.
In the following we neglect the disconnected pieces, which is expected to have a small effect, since their contribution is identically zero in the (degenerate) $N_f=3$
case (since $\sum_fq_f=0$). We note that the same choice has been made in all previous studies \cite{Gupta:2003zh,Aarts:2007wj,Ding:2010ga,Brandt:2012jc,Kaczmarek:2013dya}.
Finally, as we have shown in Sec.~\ref{sec:susc}, the contribution from 
disconnected diagrams to the charge susceptibility is
negligible.

\begin{figure}[t]
\centering
 \includegraphics[width=\textwidth]{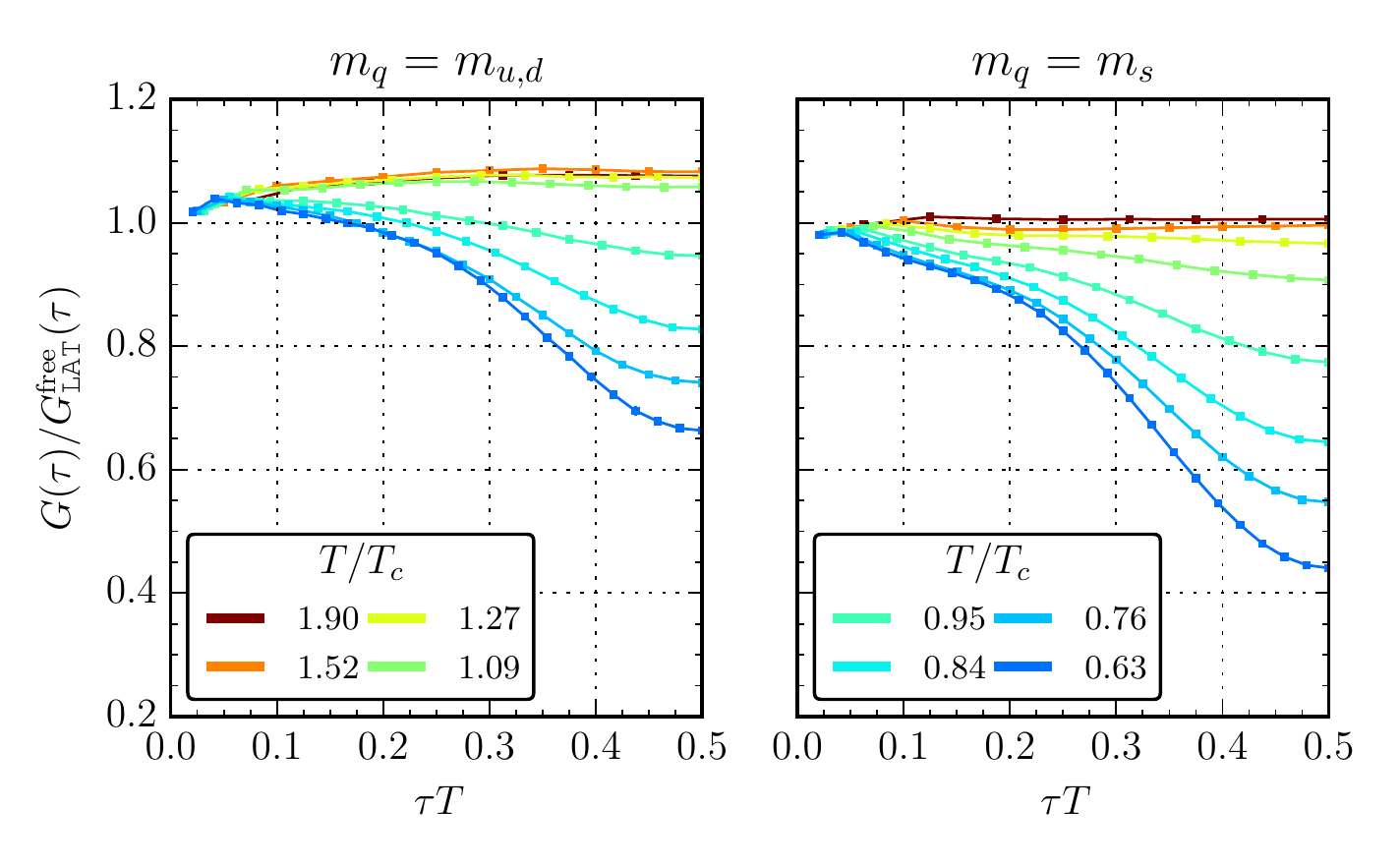}
\caption{
Conserved vector current correlator $G(\tau)$, normalised by the corresponding correlator for free lattice fermions, for light (left) and strange (right) quarks,  at different temperatures. 
}
\label{fig:consclover_lat}
\end{figure}

We have computed the conserved vector current correlator $G(\tau)=\sum_i G_{ii}(\tau)$ for the ensembles presented in Table \ref{tab:lattice-b}. In Fig.~\ref{fig:consclover} the results are shown for the light quarks (left) and strange quark (right) at all  the temperatures, as a function of Euclidean time.
Note that we present the correlators at the different temperatures as
a function of $\tau T$, which has the effect of separating them, even
when they have identical decay. The correlators are symmetric about $\tau T =1/2$.

To study the effect of increasing the temperature, we show in Fig.~\ref{fig:consclover_lat} the vector correlators normalised by the free (noninteracting) correlators on the lattice, again for both the light
and strange quarks. We observe a clear difference between the low temperature phase, where
this ratio decreases with increasing $\tau$, and the high temperature region where the ratio is relatively constant and close to unity, demonstrating that the quarks are quasi-free. The effect  of the heavier strange quark mass is clearly visible, both below and above $T_c$.
For the light quarks, we observe that all four correlators above $T_c$ follow the same pattern and exceed the free correlator by about 7\%. This may partly be due to the choice of bare parameters in the free lattice calculation, where we choose the bare anisotropy equal to the renormalised one and the bare quark mass $\hat m_0=0$. 
On the other hand, a similar enhancement above the free correlator has been observed analytically in a next-to-leading order perturbative calculation  \cite{Burnier:2014cna}. 
For the strange quark, the ratio at the highest temperatures is consistent with 1.

\begin{figure}[t]
\centering
 \includegraphics[width=0.95\textwidth]{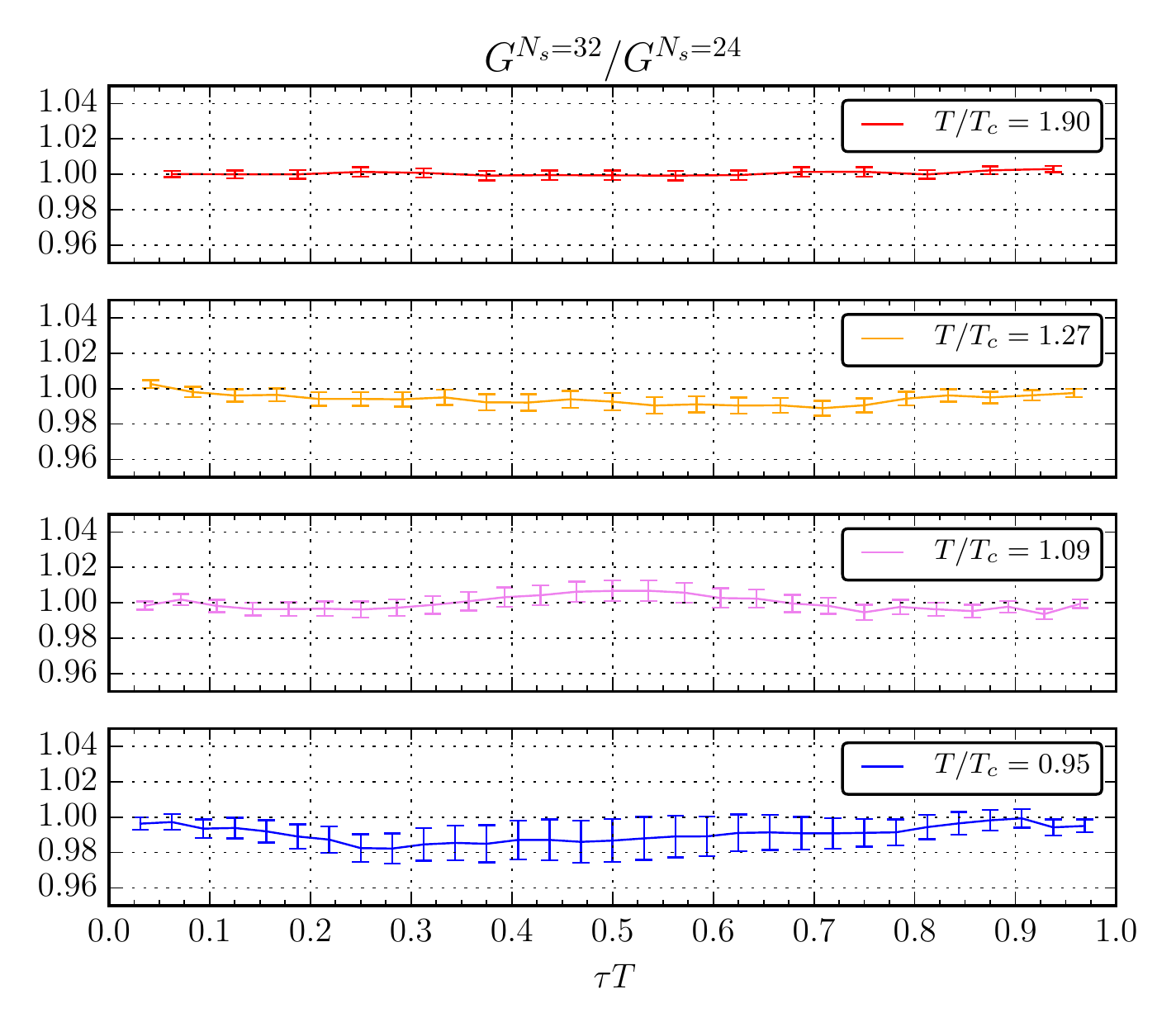}
\caption{ Estimate of finite-size effects: ratio of the conserved vector current correlator
  $G(\tau)$ for $N_s=32$ and $N_s=24$,  for four temperatures.
  }
\label{fig:consclover_vol}
\end{figure}

At four temperatures, three above and one below $T_c$, we have access to different spatial volumes, namely $L_s\sim$ 2.9 respectively 3.9 fm, or $N_s=24$ and 32. 
To study the finite-size effects, we show in Fig.~\ref{fig:consclover_vol} the ratio of the $N_s=32$ to the $N_s=24$ correlators.
We observe that  finite-size effects are at the percent level or less and decrease at higher temperature.

\subsection{Spectral functions and conductivity}
\label{subsec:specfn}

To obtain the spectral functions and conductivity from the correlators, the spectral representation 
(\ref{eq:spectr}) has to be inverted. For this we follow the same procedure as in our previous 
work \cite{Aarts:2007wj,Amato:2013naa}, namely the Maximum Entropy Method (MEM)  \cite{mem}. Other possibilities 
include the use of a physically motivated Ansatz for the spectral function, with a number
of parameters to be determined \cite{Brandt:2012jc,Ding:2010ga,Kaczmarek:2013dya}, as well as alternative inversion methods 
 \cite{Burnier:2011jq,Burnier:2012ts,Burnier:2013nla}.

Since the implementation of MEM has been presented in previous works  
\cite{mem,Aarts:2007wj,Amato:2013naa}, we summarise here only the main ingredients.
At large $\omega$ and nonzero $\tau$, the kernel $K(\tau,\omega)$ is exponentially  suppressed, hence one may impose an upper limit $\omega<\omega_{\rm max}$. The finite interval 
$0\leq\omega<\omega_{\rm max}$ is then discretised using $N_\omega$
points. Typical values are $a_\tau\om_{\rm max}=3$ and
$N_\om=1000$. 
  Eq.~\eqref{eq:spectr} has the form of a generalised Laplace
  transform, the inverse of which is known to be an ill-posed problem.
In MEM one extracts the most probable spectral function
$\rho(\omega)$, given some prior knowledge $H$ and the data $D$. This
is expressed as a conditional probability, via Bayes theorem,
\begin{equation}
  P[\rho| D H]=\frac{P[D |\rho H ]P[\rho|H ]}{P[D| H]} \propto \exp(-L + \alpha S),
\end{equation}
where $L=\frac{1}{2}\chi^2$ is the standard likelihood function and $S$ is the
Shannon-Jaynes entropy, 
\begin{equation}
\label{default}
 S=\int_0^\infty \frac{d\omega}{2\pi}\left[\rho(\omega)-m(\omega)-
    \rho(\omega)\ln\frac{\rho(\omega)}{m(\omega)}\right].
\end{equation}
Here $m(\omega)$ is the default model which implements the prior
information on $\rho(\omega)$ in the absence of the data $D$.  The
result for $\rho(\omega)$ is then obtained by extremising $P[\rho| D
H]$. To do this, we use a modification \cite{Aarts:2007wj} of Bryan's algorithm
\cite{bryan} which cures the $1/\om$ instability of the kernel $K(\tau,\om)$ at small $\omega$.
The default model we use is \cite{Aarts:2007wj}

\begin{figure}[t]
\centering
 \includegraphics[width=\textwidth]{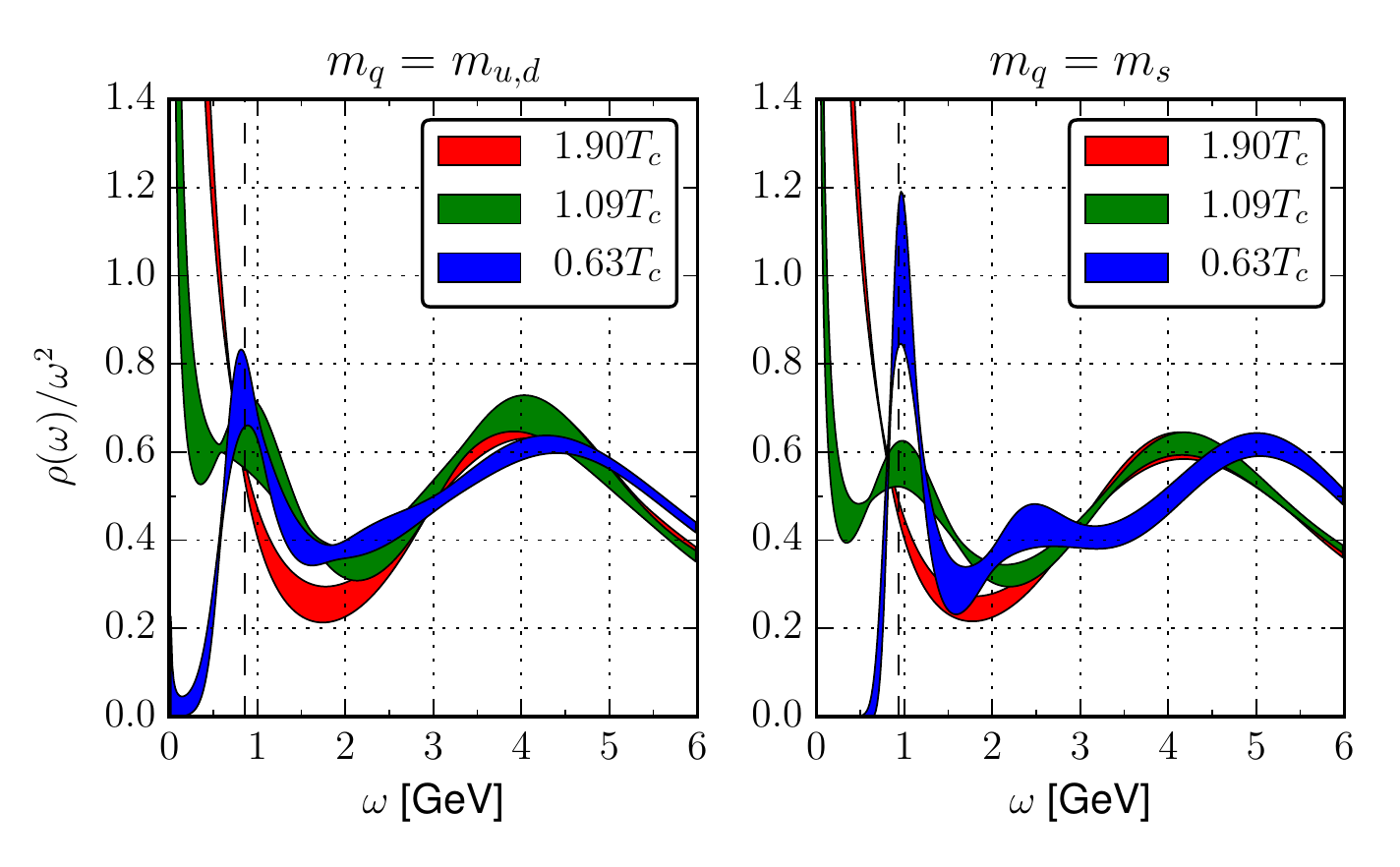}
\caption{Spectral functions $\rho(\om)/\om^2$ for light (left) and
  strange quarks (right) for three temperatures. The filled area is the statistical error
  from jackknife. The vertical dashed lines indicate the mass of the
  corresponding vector meson \cite{Lin:2008pr}.}
\label{fig:spectr_w2}
\end{figure}

\begin{equation}
m(\omega)=m_0 (b +\omega) \omega,
\label{eq:defm}
\end{equation}
where $m_0$ is an overall normalisation set by fitting the correlator
to the trial function obtained by using $m(\omega)$ in the
convolution integral (\ref{eq:spectr}). This default model is chosen because it matches the perturbative $\omega$-dependence at large $\omega$ in the continuum theory (on the lattice this behaviour is modified due to the finite Brillouin zone \cite{Karsch:2003wy,Aarts:2005hg}) and allows a nonzero
value of $\rho(\omega)/\omega$ as $\omega\to 0$ and hence a nonzero
conductivity $\sigma$ according to the Kubo relation (\ref{eq:kubo}). 
As always, it is essential to check that the spectral functions obtained
are independent of the choices made in the MEM
procedure, including the choice of default model and its parameters. This is discussed in some detail in the next subsection, after presenting the results.

\begin{figure}[t]
\centering
 \includegraphics[width=\textwidth]{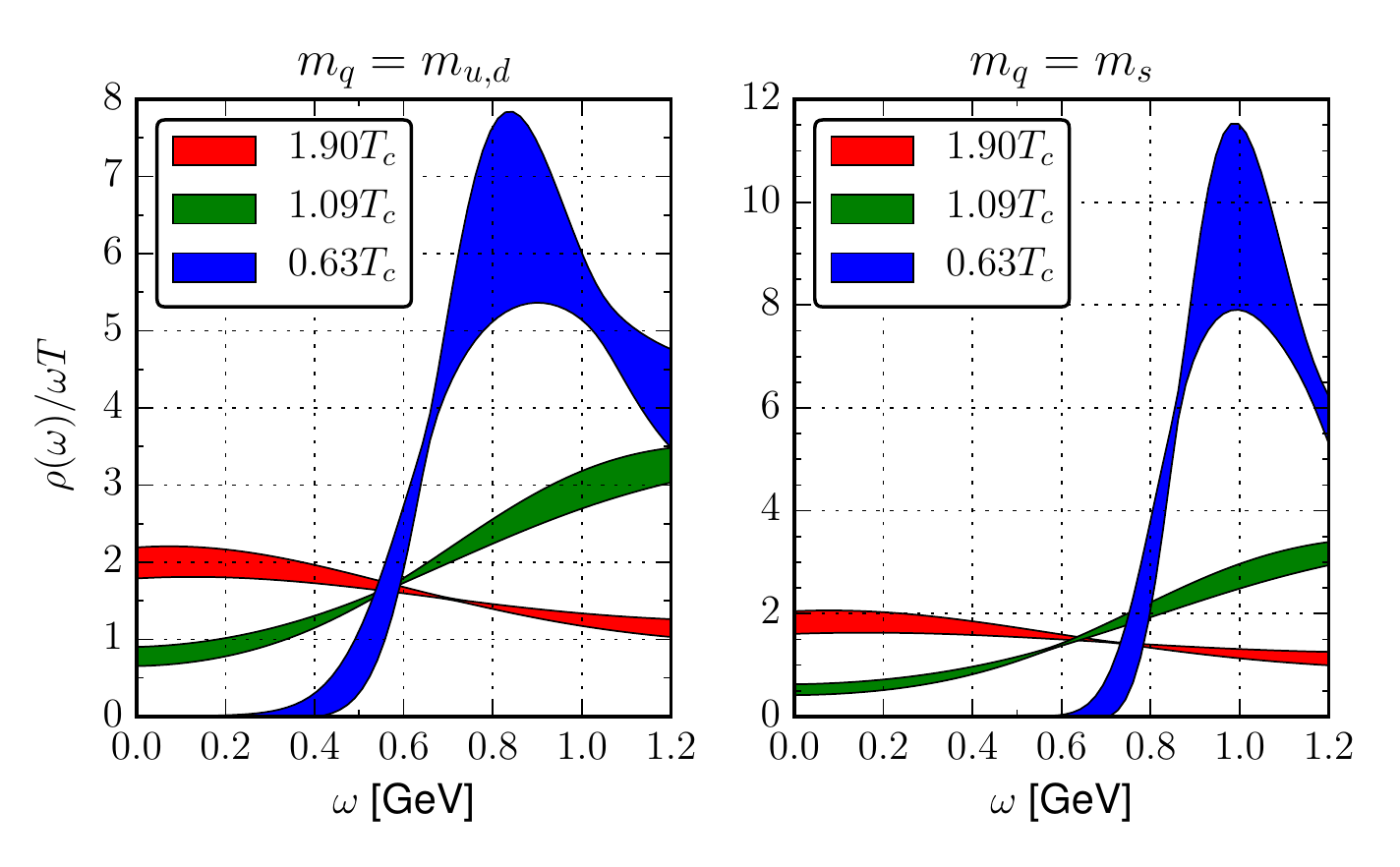}
\caption{Spectral functions $\rho(\omega)/\omega T$
  for the light (left) and the strange quarks (right) for three temperatures. The filled area
  is the statistical error from jackknife.  The intercept is
  proportional to $\sigma/T$.}
\label{fig:spectr_wT}
\end{figure}

The spectral functions obtained with MEM are shown in Fig.~\ref{fig:spectr_w2}, normalised as $\rho(\om)/\om^2$. The spectral functions are shown for light (left) and strange (right) quarks at three representative temperatures spanning the entire range. 
We always use the largest volume, $N_s=32$, when available.
The vertical dashed lines correspond to an estimate of the mass of the ground state in the corresponding vector channel at  zero temperature \cite{Lin:2008pr}.
The MEM analysis indeed indicates a peak at this value below $T_c$, which becomes less pronounced and disappears as the temperature increases.  
The divergence at small $\om$ at the higher temperatures is due to the transport peak. This is emphasised in Fig.~\ref{fig:spectr_wT} where $\rho(\omega)/\omega T$ is shown. According to the Kubo relation (\ref{eq:kubo}), the intercepts are proportional to $\sigma/T$. We observe a conductivity which is clearly nonzero above $T_c$ and which depends on the quark mass.

\begin{figure}[t]
\centering
  \includegraphics[width=0.84\textwidth]{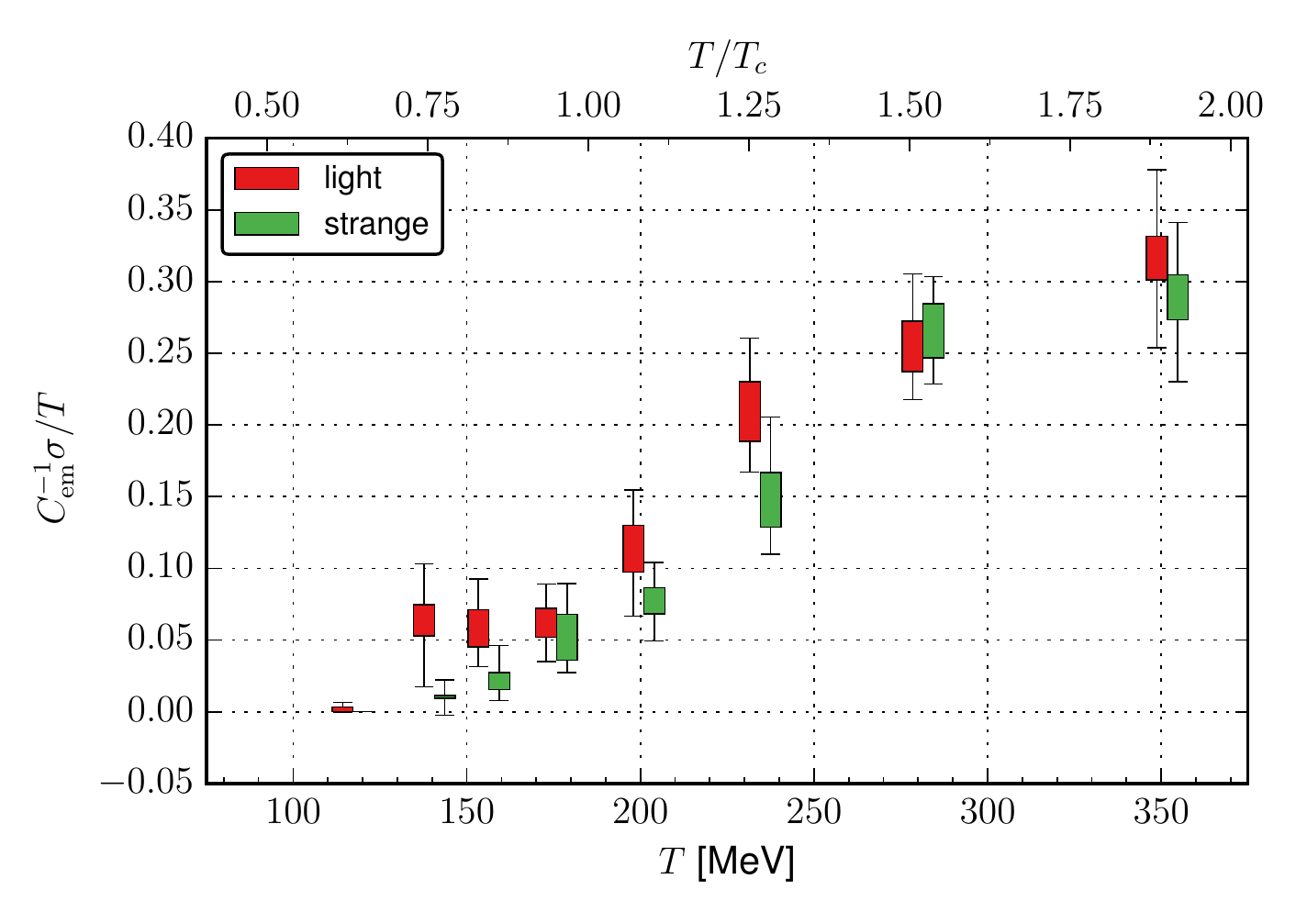}
  \includegraphics[width=0.84\textwidth]{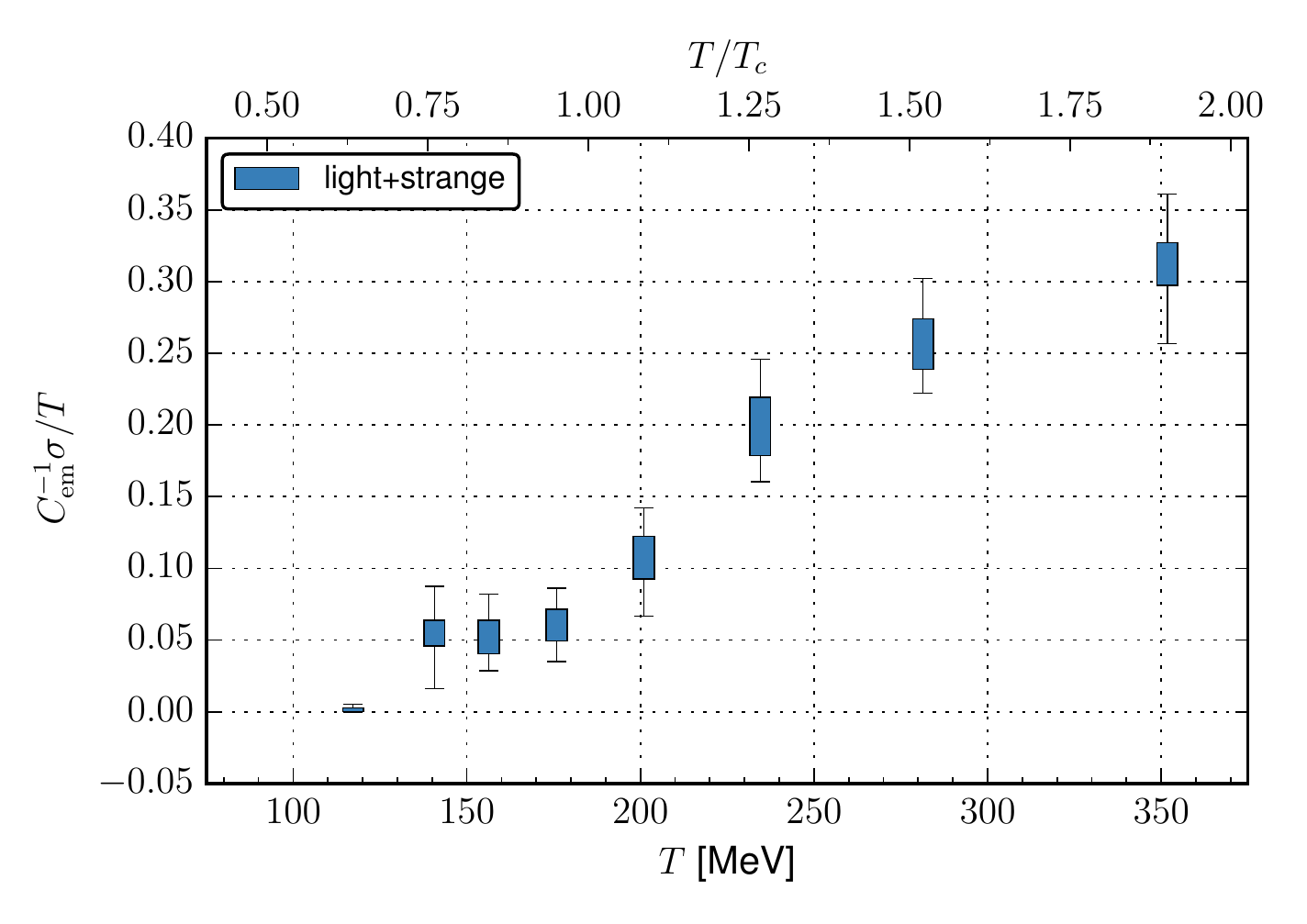}
\caption{Temperature dependence of $C_{\rm em}^{-1}\sigma/T$ for light and strange quarks separately, slightly shifted for clarity (above) and combined (below). 
The vertical size of the rectangles reflects the systematic uncertainty
  due to changes in the default model, while the whiskers depict the
  statistical jackknife error on top of this.  
    }
\label{fig:cond}
\end{figure}

The final results for the conductivity are shown in Fig.~\ref{fig:cond} as a function of the temperature.
 We present the result as $C_{\rm em}^{-1}\sigma/T$ where $C_{\rm em}$ was defined in Eq.~(\ref{eq:Cem}). 
The results are shown for the light and strange quarks separately and for all three quarks combined. 
Note that we always first construct the electromagnetic current operator with the correct weighting of the quark charges and then apply MEM to the resulting correlators.
The systematic uncertainty due to the choice of the parameter $b$ in
the default model (\ref{eq:defm}) is represented by the vertical size
of the filled rectangles. This is discussed further below. The
statistical uncertainty due to the finite number of configurations is
represented by the upper and lower whiskers emanating from the
rectangles, so that the total spread 
of values (statistical and systematic) is given by the size of the
error bars.

We observe that the contributions from the light and the strange
quarks are comparable, except in the crossover region, where the
strange quark contribution is suppressed.  Note, however, that in the total conductivity the strange quark contribution is in any case suppressed with respect to the light quarks, due to the different electromagnetic prefactors:
  $C_{\rm em}^\ell=5e^2/9$ versus  $C_{\rm em}^s=e^2/9$.

\subsection{Systematics}

\begin{figure}[t]
    \centering
 \includegraphics[width=0.8\textwidth]{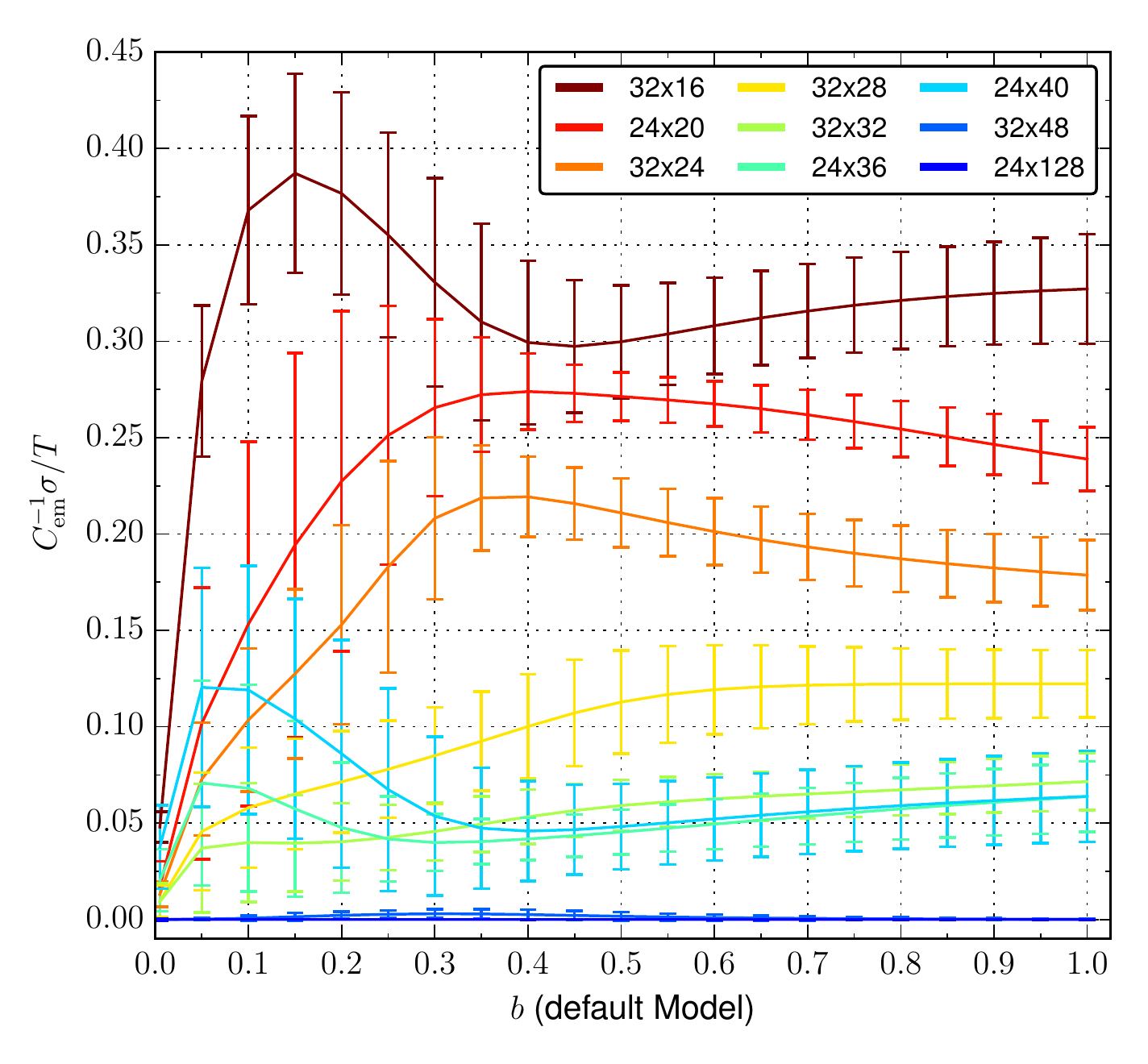}
\caption{Dependence of the conductivity on the parameter $b$ in the default 
model for the light quarks.
}
\label{fig:testb}
\end{figure}

In order to have confidence in the results, it is necessary to study the systematic uncertainties in the MEM analysis. 
As in our previous work, the sensitivity to  $\omega_{\rm max}$ is
modest, provided that $3 \lesssim a_\tau\om_{\rm max} \lesssim
5$. Here we show results from tests varying the $b$ parameter in the
default model, excluding intermediate time points, and varying the
choice of time range included in the analysis.

In Fig.~\ref{fig:testb} we show the dependence of the conductivity on
the default model parameter $b$. The results are stable provided
  $b\gtrsim 0.4$.
We therefore use the range $0.3 \leq b \leq 1.0$ to define the systematic 
error coming from the default model in our conductivity determination, 
see Fig.~\ref{fig:cond}.

\begin{figure}[t]
    \centering
 \includegraphics[width=\textwidth]{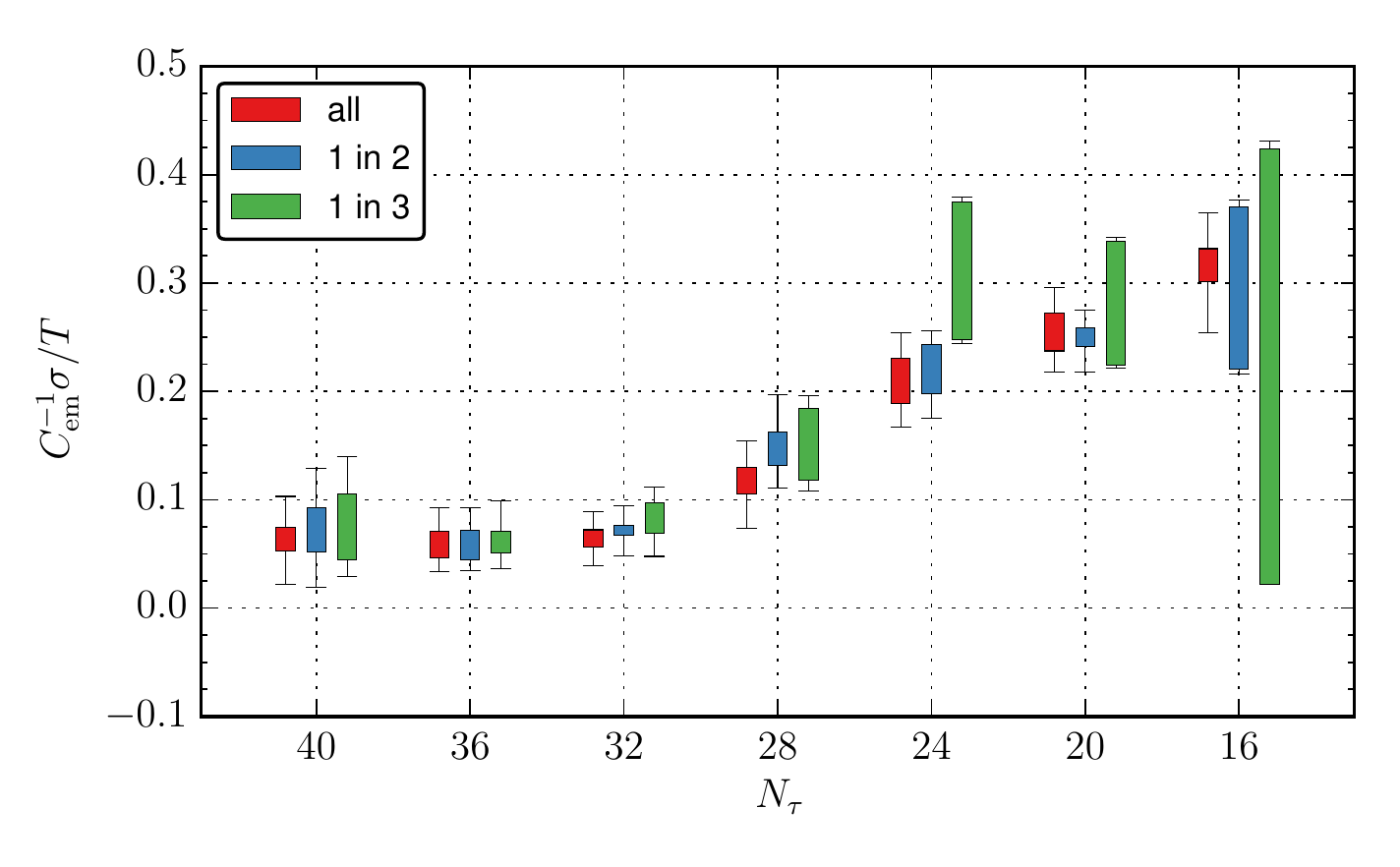}
\caption{Dependence of the conductivity on the inclusion of Euclidean time points: all, 1 in 2, or 1 in 3, always starting at $\tau=4a_\tau$, for the light quarks.
}
\label{fig:testanis}
\end{figure}

\begin{figure}[t]
    \centering
\includegraphics[width=0.8\textwidth]{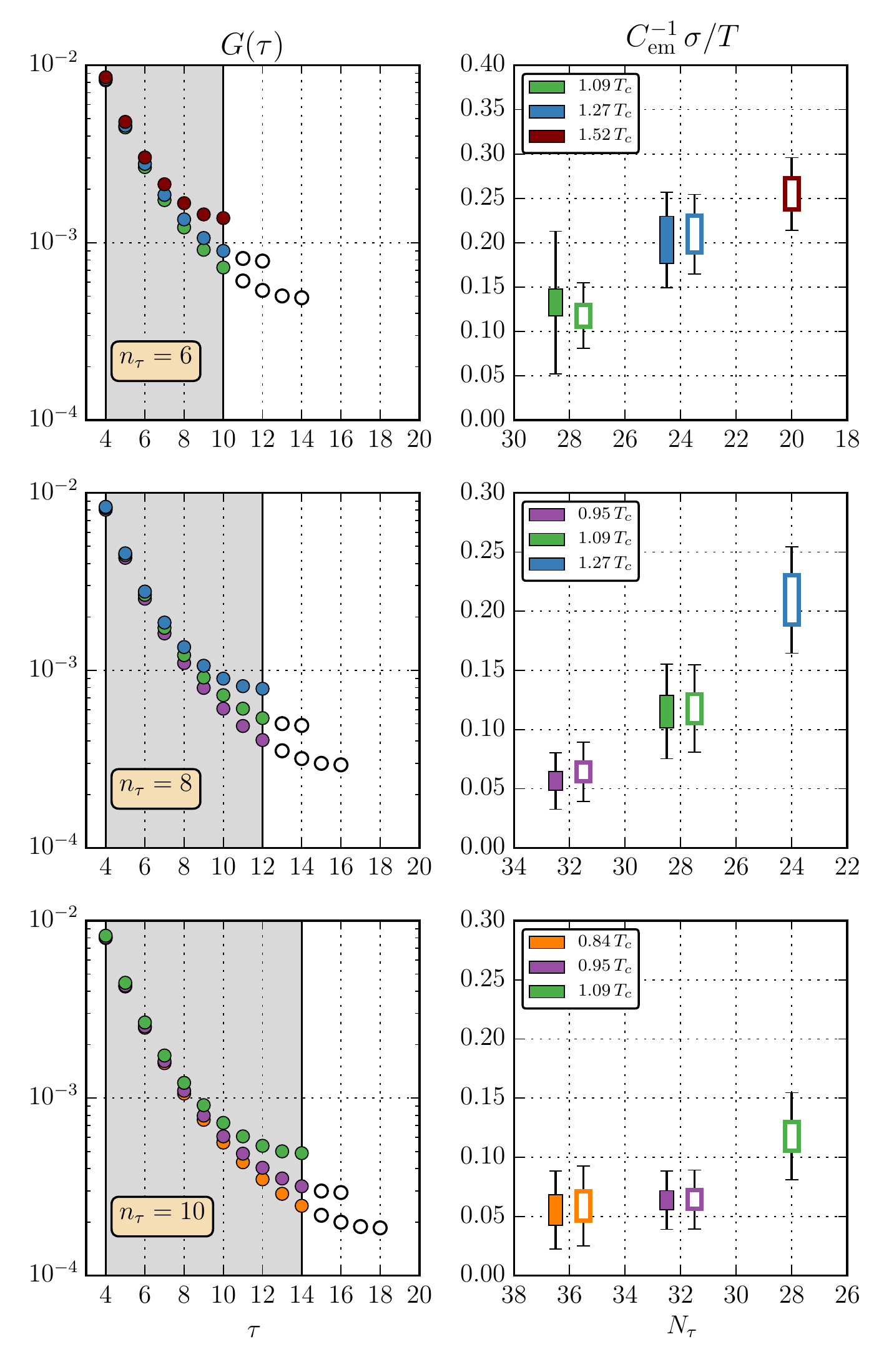}
\caption{Stability tests, for light quarks, discarding the last time slices in the
  correlators, for three sets of temperatures in each row. See main
  text for details. 
  }
\label{fig:test1}
\end{figure}

On our anisotropic lattices, the temporal lattice spacing is
smaller than the spatial one, with $\xi=a_s/a_\tau=3.5$. Hence we have
more time slices available for the MEM analysis than on an isotropic
lattice with the same spatial lattice spacing. One may question how
this improves the results, if at all. We test this by including in the
MEM analysis either all time slices (always discarding the first
four), or one in two, or one in three. With our specific value of
 $\xi$, the latter is roughly equivalent to the isotropic case.
The results are shown in 
Fig.~\ref{fig:testanis}. We observe that at the lower temperatures the results are manifestly stable and consistent and hence an isotropic lattice would suffice. On the other hand, at the higher temperatures the benefit of the anisotropy is clearly visible: while not affecting the central value substantially, it greatly reduces the systematic uncertainty in the reconstruction.  
These results indicate the robustness of the results and the necessity
of using anisotropic lattices.

Finally, we assess the uncertainty arising from the choice of  time window used in the MEM analysis.
Since we work at a fixed lattice spacing, increasing the temperature implies having fewer time slices available. 
Hence, one possibility could be that the observed temperature dependence of the conductivity is simply an artefact due to the different number of time slices available and hence not physical.
We test this by using MEM with restricted time windows at various temperatures, see Fig.~\ref{fig:test1}. 
In each row we consider three temperatures. We always perform the MEM analysis starting at $\tau/a_\tau=4$.
We then include either all time slices available (filled {\em and}
open symbols in the left plots), or constrain the number of time
slices by the highest temperature in each row (filled symbols
only). In this way we can study the effect of adding more time points
as the temperature is decreased. The resulting conductivities are
shown on the right. Here the filled
symbols are obtained using the restricted MEM analysis, while the open
symbols indicate the results with all available time slices
used. The results are remarkably stable, with agreement between open and filled symbols at each temperature, including the highest one shown. 
We hence conclude that the observed dependence of the conductivity on the temperature is a thermal effect,
rather than a bias introduced in our method.

\section{Diffusion coefficient}
\label{sec:diff}

\begin{figure}[t]
\centering
 \includegraphics[width=\textwidth]{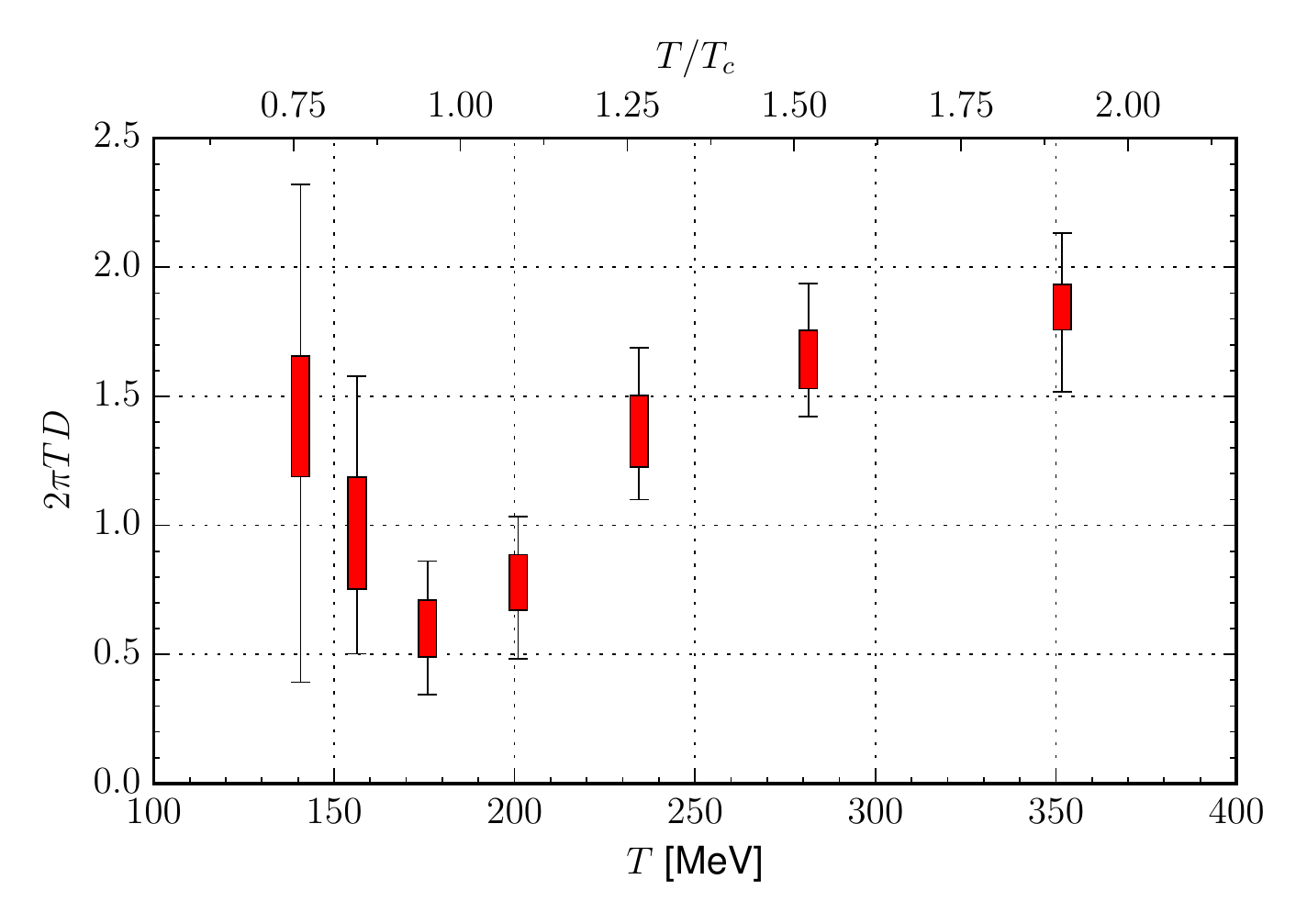}
\caption{Diffusion coefficient $D$ multiplied by $2\pi T$ as a
  function of the temperature $T$, using $D=\sigma/\chi_Q$.
  The vertical size of the rectangles represents the systematic uncertainty due to the uncertainty in the estimate of the
  conductivity (see Fig.~\ref{fig:cond}), while the whiskers indicate
  the statistical jackknife error from both $\sigma$ and $\chi_Q$. 
}
\label{fig:diffusion}
\end{figure}

We are now in the position to combine the results for the conductivity $\sigma$ and the charge susceptibility $\chi_Q$ 
to obtain the charge diffusion coefficient $D=\sigma/\chi_Q$.  This
ratio is independent of the electromagnetic factor $C_{\rm em}$. We present
the result for the dimensionless combination $2\pi TD$ in
Fig.~\ref{fig:diffusion}. The vertical size of the rectangles
represents the systematic uncertainty coming from the determination of
the conductivity, while the whiskers indicate the statistical
uncertainty in both $\sigma$ and $\chi_Q$.

The first observation we make is that the diffusion coefficient is of the order of $1/(2\pi T)$. In order to judge whether this is a sensible result, we note that in strongly coupled theories, in particular those which can be treated with holography, this is exactly the magnitude that is expected.  
 For instance, the  diffusion coefficient for $R$-charge in ${\cal N}=4$ Yang-Mills theory at nonzero temperature equals $1/(2\pi T$) \cite{Policastro:2002se,Teaney:2006nc,Son:2007vk}.
On the other hand, in weakly coupled theories the diffusion coefficient, being proportional to the mean free path, is large and diverges as the interactions are turned off. Hence our results are consistent with strongly coupled real-time dynamics in the quark-gluon plasma in this temperature range and in the hydrodynamic regime.

The second observation is a dip in the diffusion coefficient in the transition region. This can be understood as follows. We first note that at high temperature $DT$ is expected to rise, since the conductivity is expected to increase, due to the diverging mean free path at weak coupling, while the susceptibility is of the order of the Stefan-Boltzmann value. Their ratio will hence grow large.
On the low-temperature side we note that the susceptibility drops rapidly in the confined phase. On the other hand, we expect the conductivity to be nonzero, since it can be assumed that a pion gas at low temperature is a conductor rather than an insulator. Hence the ratio will again lead to a rise of $DT$. 
 This then naturally leads to a minimum around $T_c$, as in the case of the ratio of the shear viscosity to entropy density \cite{Kovtun:2004de,Csernai:2006zz}.
As a side remark we note that a successful numerical evaluation at very low temperature along the lines followed here will be unlikely, since $D$ involves the ratio of two suppressed quantities in the confined phase.

We note here that a plot similar to Fig.~\ref{fig:diffusion} was constructed in Ref.~\cite{Ling:2013ksb}, by combining the conductivity results for the two light flavours from Ref.~\cite{Amato:2013naa} with the (continuum-extrapolated) susceptibility results from Ref.~\cite{Borsanyi:2011sw}.
The conductivity and (quark number) susceptibility have also been computed in Ref.\ \cite{Ding:2010ga} for quenched QCD and in 
Ref.\ \cite{Brandt:2012jc} for QCD with $N_f=2$ flavours, but the resulting diffusion coefficient was not given. Note that the latter also contains a comparison with ${\cal N}=4$ Yang-Mills theory.

Finally, we remark that an attempt to determine the charm diffusion coefficient can be found in Ref.~\cite{Ding:2012sp} using quenched lattice simulations on large and fine isotropic lattices, with the finding that  $D\sim 1/(\pi T)$ in the deconfined phase.
For very heavy quarks various diffusion coefficients are being determined using heavy-quark effective theory, see e.g.\
Refs.~\cite{CaronHuot:2009uh,Banerjee:2011ra,Francis:2013cva} and references therein.

\section{Conclusion}
\label{sec:conclude}

The main result in this paper is the determination of the
electrical conductivity and charge diffusion coefficient at
nonzero temperature in  QCD with $N_f=2+1$ quark flavours, using
anisotropic lattice QCD simulations. 

Our results for the conductivity $\sigma$ confirm our previous findings
  where only the $u$ and $d$ quark contributions were taken into
  account: $\sigma/T$ increases by a factor of 5--6 in our temperature
range, which spans the chiral and deconfinement transition.  We note
that the results for the conductivity at the lowest
temperature should be treated with caution, since a possible narrow
transport peak resulting from hadronic interactions would not be
detectable with our methods.
We find that the diffusion coefficient is of the order of $1/(2\pi T)$ and has a dip around the transition temperature between the confined and the deconfined phase. This is consistent with a strongly-coupled quark-gluon plasma in the hydrodynamic regime.

In order to reach this result, we have used the Maximum Entropy Method to construct spectral functions from the numerically determined Euclidean correlators of the conserved vector current. The conductivity then follows from the linear behaviour of the spectral functions at small energies. Independently we have determined various second-order susceptibilities and found agreement with previous results. The diffusion coefficient is given by the ratio of the electrical conductivity to the charge susceptibility.

As an outlook, we note that there are various things that can be improved. Besides MEM, it might be useful to apply other recently developed inversion methods \cite{Burnier:2011jq,Burnier:2012ts,Burnier:2013nla}. It should be stated that our results are robust against variation of several systematic input variables, most notably those related to the Euclidean time interval and number of Euclidean time points included. Here we found that the anisotropy is essential at the highest temperatures.
Concerning our ensembles, we note that the spatial lattice is relatively coarse and that the light quarks are heavier than in nature, which affects the transition temperature. Hence it is worthwhile to repeat the analysis with lighter quarks on finer lattices. An increase of the anisotropy on the other hand will allow for even better control on systematics of the inversion. We hope to address some of these issues in the future.

\section*{Acknowledgements}

We thank Benjamin J\"ager and Prem Kumar for discussion. 
We are grateful to the Hadron Spectrum Collaboration for providing the zero-temperature ensemble. 
For computational resources we thank the STFC funded DiRAC Facility,  HPC Wales,  PRACE via grants 2011040469 and 2012061129, and ICHEC.
GA is supported by STFC, the Royal Society, the Wolfson Foundation and the Leverhulme Trust. 
CRA is supported by STFC and the Leverhulme Trust. He thanks the Institute for Nuclear Theory at the University of
Washington for its hospitality and the Department of Energy for partial support during the completion of this work.
AA was supported by a Swansea University postgraduate scholarship and acknowledges  European Union Grant Agreement number 238353 (ITN STRONGnet) and Academy of Finland grant 1267286  for support  during the completion of this work.
SH is supported by STFC.

\end{document}